\documentclass[twocolumn,superscriptaddress]{revtex4-2}
\pdfoutput=1
\usepackage{algorithm2e}
\usepackage{amsmath}
\usepackage{amssymb} 
\usepackage{braket} 
\usepackage[utf8]{inputenc}
\usepackage{cancel} 
\usepackage{mathtools}
\usepackage{mathbbol}    
\usepackage[svgnames]{xcolor}
\usepackage[colorlinks=false,hidelinks]{hyperref}
\usepackage{siunitx}
\usepackage[pdftex]{graphicx}
\usepackage[toc,acronym]{glossaries} 

\begin{document}
\RestyleAlgo{ruled}
\title{Exclusive-or encoded algebraic structure for efficient quantum dynamics}
\author{Lukas Broers}
\affiliation{Center for Optical Quantum Technologies, University of Hamburg, 22761 Hamburg, Germany}
\affiliation{Institute for Quantum Physics, University of Hamburg, 22761 Hamburg, Germany}
\author{Ludwig Mathey}
\affiliation{Center for Optical Quantum Technologies, University of Hamburg, 22761 Hamburg, Germany}
\affiliation{Institute for Quantum Physics, University of Hamburg, 22761 Hamburg, Germany}
\affiliation{The Hamburg Center for Ultrafast Imaging, 22761 Hamburg, Germany}

\begin{abstract} 
We propose a formalism that captures the algebraic structure of many-body two-level quantum systems, and directly motivates an efficient numerical method. 
This formalism is based on the binary representation of the enumeration-indices of the elements of the corresponding Lie algebra.
The action of arbitrarily large elements of that algebra reduces to a few bit-wise exclusive-or operations.
This formalism naturally produces sparse representations of many-body density operators, the size of which we control through a dynamic truncation method.
We demonstrate how this formalism applies to real-time evolution, dissipative Lindblad action, imaginary-time evolution, and projective measurement processes.
We find that this approach to calculating quantum dynamics scales close to linearly with the number of non-zero components in the density operator. 
We refer to this exclusive-or represented quantum algebra as ORQA\@. 
As a proof of concept, we provide a numerical demonstration of this formalism by simulating quantum annealing processes for the maximum independent set problem for up to 22 two-level systems.
\end{abstract}

\maketitle

\section{Introduction}
\begin{figure*}[t]
\centering  
\includegraphics[width=0.85\linewidth,trim={0 0 0 7.9cm},clip]{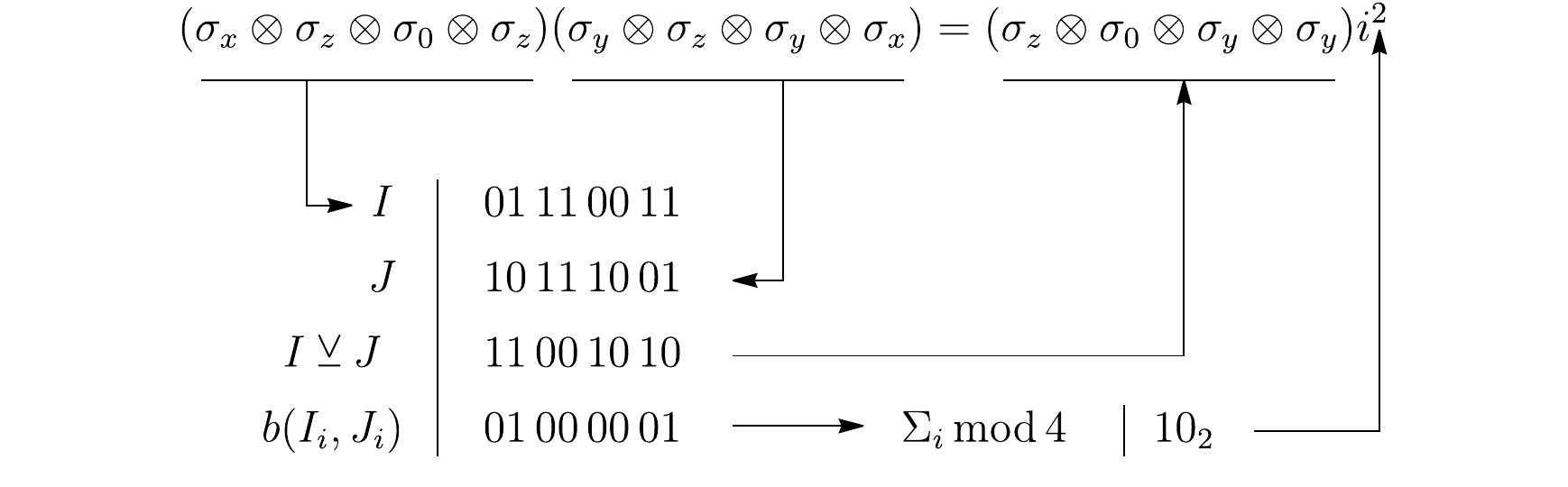} 
\caption{
    \textbf{Illustrative example of the exclusive-or ($\veebar$) structure of $\mathfrak{su}(2^n)$.}
    The indices of tensor-products (in this example $n=4$) of Pauli matrices when encoded as binary strings $I$ and $J$, reproduce the group structure through the exclusive-or ($\veebar$) operation without the need for an explicit exponentially large matrix-representation.
    The structure factor is encoded in the function $b(j,k)$ as in Eq.~\ref{jxork} and Tbl.~\ref{table}.
}\label{illus}
\end{figure*}

It is difficult to overstate the relevance of composite two-level systems, and the dynamics thereof, for a vast number of quantum many-body physics.
A prime example is quantum information theory~\cite{Watrous18} and quantum computing~\cite{NielsenChuang10} in particular, where information is encoded in sets of interacting two-level systems. 
Similarly, quantum optics~\cite{gross82,temnov05,garraway11,Browne17} and various spin systems~\cite{Affleck89,Khaneja01,Backens19}, are fields of study that heavily rely on the structure of composite two-level systems.
Analytical transformations like those put forth and named after Jordan-Wigner~\cite{Jordan28}, Schwinger~\cite{schwinger52}, and Holstein-Primakoff~\cite{Holstein40} are kept in high regard as they map second quantization operators onto spin-structures.
In the field of quantum simulation~\cite{Georgescu14,Altman21}, the particular method of quantum annealing~\cite{Finnila94,Hauke20} fundamentally relies on mapping classically hard problems onto Ising-like~\cite{Biamonte08,Ising25} spin models~\cite{lucas14}, through methods like quadratic unconstrained binary optimization (QUBO)~\cite{glover19}. 
Variational closed-loop optimization on quantum devices has similarly been used to treat combinatorics problems~\cite{Ebadi22}.
As such, the recent exploration of quantum machine learning~\cite{farhi14,Biamonte17,Zhou20} heuristics deepens the connection between quantum physics and computer science methodologies.

Efficient computational methods are invaluable in many branches of research of modern physics, and other natural sciences alike.
Many-body quantum systems, composed of two-level systems or not, are notorious due to their exponential complexity, which at the same time is the potential that quantum computers promise to harness~\cite{Feynman82}.
Many numerical methods have been established to approximate and accelerate calculations of this highly complex class of quantum systems.
Approaches such as the density-matrix renormalization group~\cite{White92,white93,DMRG}, tensor networks~\cite{orus14,orus19,Cirac21}, and quantum Monte-Carlo methods~\cite{Ceperley86,Austin12} are staples of the contemporary quantum physicist's toolbox and make it possible to study these challenging systems through reduced representations and adaptive means~\cite{zwolak94,Daley04,verstraete04}.

In this work, we present an efficient description of the action of elements of the Lie algebra $\mathfrak{su}(2^n)$, which captures the structure of $n$-body composites of two-level systems.
This method avoids an explicit representation and infers all calculations from bit-wise operations on binary-indices that enumerate the elements of $\mathfrak{su}(2^n)$.
This approach reduces the overhead of exponentially large matrix products to a few binary operations consisting primarily of the exclusive-or ($\veebar$).
We refer to this formalism as the exclusive-or represented quantum algebra (ORQA).
From this formalism we directly infer an efficient and natural numerical approach for calculating the real-time evolution, dissipative Lindblad action, imaginary-time evolution, and projective measurements in $n$-body composites of two-level quantum systems.
While the exponential scaling of quantum complexity generally applies, ORQA uses a sparse representation of density operators, in which the numerical effort scales linearly with the number of non-zero components in the density operator rather than with the naive size of the Hilbert-space.
Therefore, we introduce a dynamic truncation method that drastically reduces the effective size of the density operator, while approximating the exact solutions well.
We numerically demonstrate the ORQA formalism in the example of quantum annealing for the maximum independent set (MIS) problem for up to 22 spins.
We propose ORQA as a general and well-performing framework for quantum dynamics, and as an addition to the numerical toolbox of quantum many-body physics.
This formalism can be used in conjunction with established numerical many-body methods and is highly parallelizable, such that we expect the performance to drastically improve further in the future. 

This work is structured as follows.
In Section~\ref{method}, we introduce the ORQA formalism analytically, highlighting its binary nature and how this motivates our computational method.
In Section~\ref{numdemo}, we present a numerical demonstration on examples of quantum annealing. 
In Section~\ref{conclusion}, we conclude our results and present possible future directions for the ORQA formalism.

\section{Methodology}
\label{method}

\begin{table}[b]
    \begin{center}
\begin{tabular}{c|cccc} 
$\,{}_j\,\,{}^k$&11&10&01&00\\
\hline
11&{00}&{11}&{01}&{00}\\
10&{01}&{00}&{11}&{00}\\
01&{11}&{01}&{00}&{00}\\
00&{00}&{00}&{00}&{00}
\end{tabular}
\end{center}
\caption{
    The binary table of the exponent $b(j,k)$ of the structure factor $i^{b(j,k)}$ of $\mathfrak{su}(2)$. 
    Concatenating the rows of this table provides the binary representation of $M$ in Eq.~\ref{magicnumber}.}\label{table}
\end{table}

\begin{figure*}[t]
\centering 
\hspace{1cm}\includegraphics[width=0.55\linewidth]{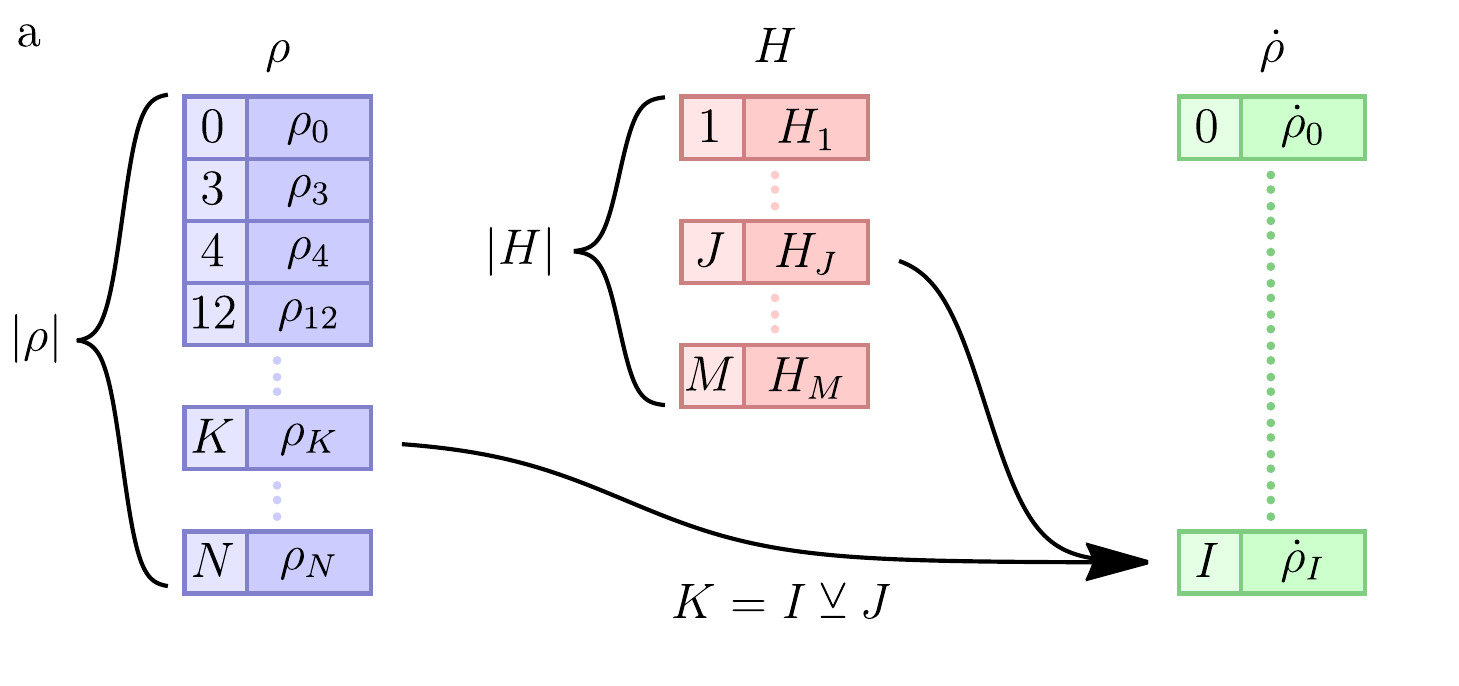}\newline
\includegraphics[width=0.32\linewidth]{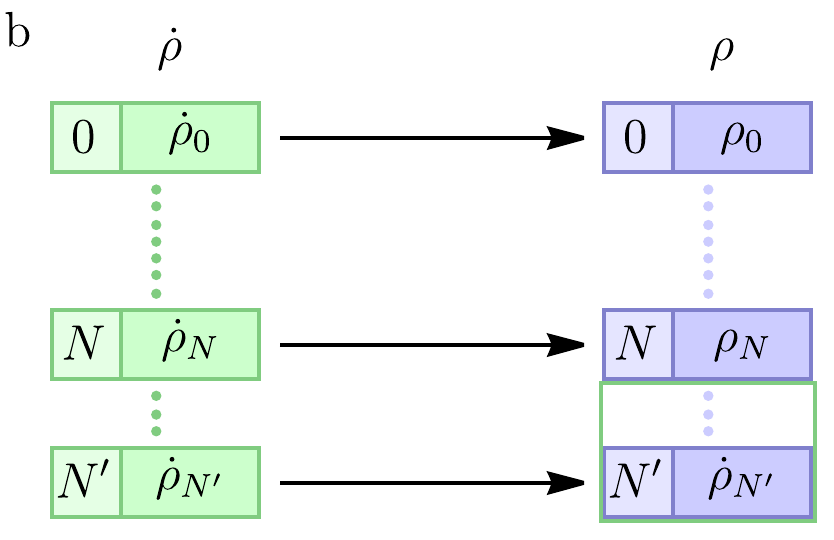}
\hspace{2cm}
\includegraphics[width=0.32\linewidth]{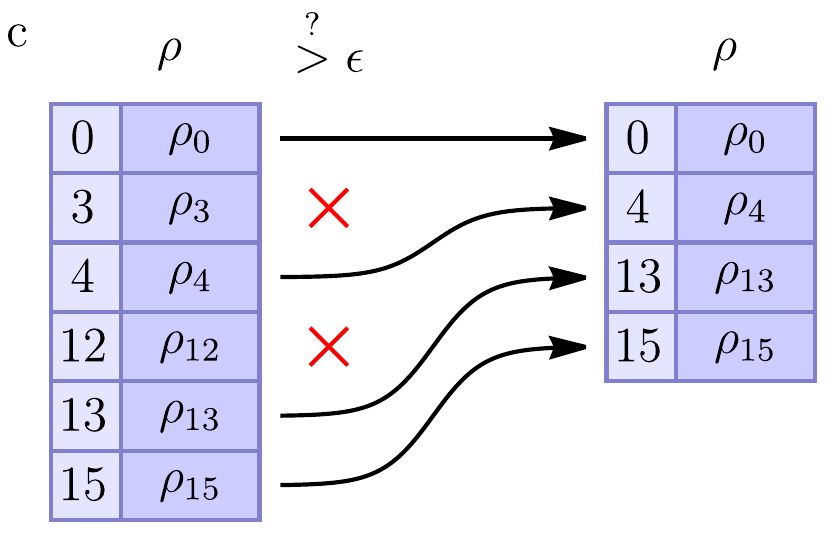}
\caption{
\textbf{Illustration of the numerical ORQA structure for quantum dynamics. }
The density operator $\rho$, its derivative $\dot\rho$, and the Hamiltonian $H$ are numerically expressed as dynamically sized maps of multi-indices $I$ and the corresponding components $\rho_I$.
Panel~(a) shows an example of the memory access of an update operation. 
Calculating a component $\dot\rho_I$ involves accessing the components $\rho_{I\veebar J}$ and $H_J$, for all $J$ for which $H_J\not = 0$.
Panel~(b) shows the one-to-one update of $\rho$ given by $\dot\rho$. 
In the case that $|\dot\rho|>|\rho|$, the size of $\rho$ is dynamically increased to include the new components.
Panel~(c) shows the truncation of $\rho$ using a threshold $\epsilon>0$. 
In the case that $|\rho_I|\leq\epsilon$, the corresponding component of $\rho$ is removed, dynamically reducing the size of $\rho$.
}\label{cartoon}
\end{figure*}
 
The dynamics of two-level quantum systems are canonically represented using the Pauli matrices 
\begin{align}
    \sigma_x&=\begin{pmatrix}0 & 1 \\ 1 & 0\end{pmatrix}&
    \sigma_y&=\begin{pmatrix}0 & -i \\ i & 0\end{pmatrix}&
    \sigma_z&=\begin{pmatrix}1 & 0 \\ 0 & -1\end{pmatrix}.
\end{align} 
These matrices are a specific choice of a representation of the Lie algebra $\mathfrak{su}(2)$.
We introduce a particular notation of the Pauli matrices using the binary representation of the subscripts.
We define
\begin{align}
    \sigma_{00} &= \mathbb{1}&
    \sigma_{01} &= \sigma_x&
    \sigma_{10} &= \sigma_y&
    \sigma_{11} &= \sigma_z.
\end{align}
This notation allows us to simplify the representation-independent extended (including the identity) algebraic structure as
\begin{align}
    \sigma_j\sigma_k &= \delta_{j,0}\sigma_k + \delta_{k,0}\sigma_j + (\delta_{j,k} - 2 \delta_{j,0}\delta_{k,0})\sigma_0 + i \epsilon^{jkl} \sigma_{l}\label{jxork1}\\
    &= i^{b(j,k)} \sigma_{j\veebar k}.\label{jxork}
\end{align}
Here $\veebar $ is the bit-wise exclusive-or operation and the function $b(j,k)$ generalizes the possible prefactors in Eq.~\ref{jxork1}. 
$\epsilon^{jkl}$ is the Levi-Cevita symbol.
We continue to write the commutator of such elements as
\begin{equation}
    [\sigma_j,\sigma_k] = (i^{b({j,k})}-i^{-b({j,k})})\sigma_{j\veebar k} = 2 i y^{j,k} \sigma_{j\veebar k},
\end{equation}
where $y^{j,k}=\mathrm{Im}[i^{b({j,k})}]\in\{1,0,-1\}$.
Analogously, the anti-commutator reads
\begin{equation}
    \{\sigma_j,\sigma_k\} = (i^{b({j,k})}+i^{-b({j,k})})\sigma_{j\veebar k} = 2 x^{j,k} \sigma_{j\veebar k}
\end{equation}
with $x^{j,k}=\mathrm{Re}[i^{b({j,k})}]$.

These factors $x^{j,k}$ and $y^{j,k}$ can in principle be computed in many different ways. 
We emphasize one method that is numerically efficient and consists of encoding $b(j,k)$ in a look-up table to minimize overhead. 
For that purpose, we collect all the possible values of $b(j,k)$ in a $4\times 4$ table that we show in Tbl.~\ref{table}.
We concatenate the rows of values of this table into a single 32-bit-string.
This bit-string provides the number 
\begin{align}
    M &= {00110100}{01001100}{11010000}{00000000}_2\nonumber\\
    &= 877449216,\label{magicnumber}
\end{align}
which hence encodes the structure factor of $\mathfrak{su}(2)$.
Reading the $(j+4k)$th two-bit string of $M$ gives $b(j,k)$.

This representation of the algebraic structure and using $M$ to calculate the structure factor is advantageous, because this replaces products of explicit representations of the Lie group elements with simple and fast bit-wise operations that avoid an explicit representation.
This is particularly advantageous moving forward, as this approach generalizes in a straight-forward manner to arbitrarily large Pauli-strings.

Consider an arbitrary tensor product of Pauli matrices $\sigma_{I_i}$ with 2-bit-strings $I_i$, for which we introduce the notation
\begin{equation}
    \otimes_{i=0}^{n-1} \sigma_{I_i} = \sigma_{I_{n-1}|I_{n-2}|\dots|I_0} = \sigma_I.
\end{equation}
Here $I_i|I_j$ denotes the direct concatenation of bit-strings such that $I$ is a $2n$-bit string.
$I$ is therefore a binary representation of a non-negative integer up to the value $4^n-1$, which allows us to enumerate all $4^n$ elements of $\mathfrak{su}(2^n)$, in addition to the identity. 
We call $I$ a multi-index.
The algebraic structure of $\mathfrak{su}(2^n)$ emerges in this formalism analogously to Eq.~\ref{jxork}.
We write
\begin{align}
    \sigma_I\sigma_J&= \otimes_{i=0}^{n-1} \sigma_{I_i}\sigma_{J_i}\\
    &= i^{\sum_{i=0}^{n-1} b(I_i,J_i)}\otimes_{i=0}^{n-1} \sigma_{I_i\veebar J_i}\\
    &= i^{B(I,J)}\sigma_{I\veebar J},
\end{align}
where we have used that the exclusive or is a bit-wise operation, i.e. $I_i\veebar J_i={(I\veebar J)}_i$, and we have introduced $B(I,J)=\sum_{i=0}^{n-1} b(I_i,J_i)$. 
Further, the commutator and anti-commutator of arbitrary Pauli-strings are, analogously to the above, written as
\begin{align}
    [\sigma_I,\sigma_J] 
    & = (i^{B(I,J)}-i^{-B(I,J)})\sigma_{I\veebar J}\\
    & = 2 i Y^{I,J} \sigma_{I\veebar J}
\end{align}
and
\begin{align}
    \{\sigma_I,\sigma_J\}
    & = (i^{B(I,J)}+i^{-B(I,J)})\sigma_{I\veebar J}\\
    & = 2 X^{I,J} \sigma_{I\veebar J},\label{anticom}
\end{align}
where $X^{I,J}=\mathrm{Re}[i^{B(I,J)}]$ and $Y^{I,J}=\mathrm{Im}[i^{B(I,J)}]$.
Note that in the calculation of these factors, only the two least significant bits of $B(I,J)$ are relevant, as any higher bit represents a multiple of four which leaves $i^{B(J,K)}$ invariant.
We show the evaluation table of $X^{J,K}$ and $Y^{J,K}$ in Tbl.~\ref{table2}.

This structure provides a formalism in which the group action of elements of $\mathfrak{su}(2^n)$ is inferred directly from the multi-indices alone, which enumerate the group elements.
This fully reduces calculating the commutator of $2^n$-dimensional representations of $n$-fold tensor products of Pauli matrices to a few simple bit-wise operations.
We illustrate this structure in Fig.~\ref{illus}.
We propose to use this to implement efficient numerical simulations of various quantum systems, with two examples discussed below.
We refer to this formalism as the exclusive-or represented quantum algebra (ORQA).

In the following sections, we show how this formalism extends efficiently to multiplications of large linear combinations of Pauli-strings, such as arbitrary $2^n$-dimensional hermitian operators.
From this we identify efficient representations of quantum dynamics described by the von Neumann equation, the Lindblad master equation, imaginary-time evolution, and projective measurement processes.
In this manner, ORQA can be utilized to efficiently, and with minimal overhead, simulate quantum physical systems that display the $\mathfrak{su}(2^n)$ structure, i.e.\ composite two-level quantum systems.

For this purpose, we write density operators $\rho$ and Hamiltonians $H$ as linear combinations such as 
\begin{align}
    \rho &= \sum_{I=0}^{4^n-1}\rho_I\sigma_I\\
    H &= \sum_{I=0}^{4^n-1}H_I\sigma_I,
\end{align} 
with $\rho_I,H_I \in \mathbb{R}$.
We note that this is a type of superoperator formalism that linearizes the density matrix formalism.
However, the exclusive-or based algebraic structure as a way to manipulate such objects has to our knowledge not been explored before.
We outline the algorithmic implementation of ORQA in App.~\ref{app-algo}.
Note that this formalism can also be applied to unitary transformations $U$ in order to study for instance the full unitary time evolution or products of unitaries, e.g.\ for the purpose of calculating the action of quantum circuits.

\begin{table}
    \begin{center}
\begin{tabular}{c|c|c} 
$B(J,K)\mod4$&$X^{J,K}$&$Y^{J,K}$\\
\hline 
00&1&0\\
01&0&1\\
10&-1&0\\
11&0&-1\\
\end{tabular}
\end{center}
\caption{
The binary table of the $\mathfrak{su}(2^n)$ structure factor components $X^{J,K}$ and $Y^{J,K}$ for commutators and anti-commutators of Pauli-strings, respectively. 
}\label{table2}
\end{table}

\subsection{Real-Time Evolution}\label{RTsection}

In the density operator formalism, the unitary time evolution of closed quantum systems is governed by the von Neumann equation 
\begin{equation}
    \dot\rho = i [\rho, H],
\end{equation}
which in the ORQA formalism we write as
\begin{align}
    \sum_{I=0}^{4^n-1} \dot\rho_I \sigma_I &= i [\sum_{I=0}^{4^n-1} \rho_I \sigma_I, \sum_{J=0}^{4^n-1} H_J \sigma_J]\\
    &= 2\sum_{I,J=0}^{4^n-1} \rho_I H_J Y^{I,J} \sigma_{I\veebar J}.\label{reprhofirs}
\end{align}
Note at this point, that the set of all $2n$-bit strings ${\{0,1\}}^{2n}$ transforms into itself under applying the exclusive-or operation $\veebar$ to all elements of that set with any one element of that set. 
That means it is 
\begin{equation}
    \{ x \veebar y, x\in {\{0,1\}}^{2n} \} = {\{0,1\}}^{2n}, \,\forall y\in{\{0,1\}}^{2n}.
\end{equation} 
Therefore, in the sum over all $4^n$ multi-indices in Eq.~\ref{reprhofirs}, we replace the multi-index $J$ with $J \veebar I$, which is simply a useful reordering of the summation.
We further invoke that $I\veebar J\veebar J~=~I$ and arrive at the component $\dot\rho_I$ of the von Neumann equation
\begin{align} 
    \dot\rho_I &= 2{\sum_{J}}^\prime \rho_{I\veebar J} H_J Y^{I\veebar J,J}\label{repH}\\
    &= 2{\sum_{J}}^\prime \rho_{J} H_{J\veebar I} Y^{J,J\veebar I}.\label{reprho}
\end{align}
The prime denotes that a sum only needs to be performed over multi-indices for which $H_J$ in Eq.~\ref{repH}, or $\rho_J$ in Eq.~\ref{reprho}, are zero, which may be known \textit{a priori}.
These representations are both correct, but they differ in whether the sum is performed over the multi-indices that capture $H$ or those that capture $\rho$.
Usually the Hamiltonian $H$ is sparse in this representation, while $\rho$ is generally not, such that Eq.~\ref{repH} may in general be numerically favorable.
We illustrate the dynamical data structure of calculating $\dot\rho_I$ in Fig.~\ref{cartoon}~(a), and updating $\rho$ with $\dot\rho$ in Fig.~\ref{cartoon}~(b).

Solving for $\dot\rho$ requires iterating over all $\rho_I$ and $H_J$, calculating $Y^{I \veebar J,J}$, and accessing $\rho_{I \veebar J}$.
This procedure approximately scales as $\mathcal{O}(|\rho| |H|)$, where $|\rho|$ and $|H|$ are the numbers of non-zero components in $\rho$ and $H$, respectively.
We omit the scaling of identifying the components $\rho_{I\veebar J}$, which is potentially constant in time, depending on the exact numerical implementation, e.g.\ using a hash map to encode $\rho$ and $H$.
Note that it is also possible to directly encode $\rho_I$ at a physical memory-address $I$ which removes any overhead from identifying $\rho_{I \veebar J}$, as one can immediately access the memory-address $I\veebar J$.
This fully reduces the calculation of $\dot\rho$ to systematically accessing pointers in memory and guarantees the scaling of $\mathcal{O}(|\rho||H|)$, however only at the expense of exponentially large memory, as all $4^n$ addresses need to be readily available, which is not tractable for large calculations.
Regardless, in the worst case it is $|\rho|=4^n$, which reproduces the exponential scaling that is expected from quantum dynamics. 
We emphasize that Eq.~\ref{reprho} for different $I$ can be parallelized well, as all $\dot\rho_I$ can be evaluated independently of each other which provides the potential for drastically increased performance.

Note that besides the von Neumann equation, the derivation in this section equally applies to the Heisenberg equation which describes the time evolution of hermitian operators
\begin{equation}
    O = \sum_{I=0}^{4^n-1} O_I \sigma_I.
\end{equation}
In some instances, the dynamic size $|O|$ might lead to favorable performance.

\subsection{Lindblad Dissipation}\label{lindsection}
In every quantum system it is useful to consider the presence of dissipative processes, which lead to modified dynamics that do not preserve the unitarity or information of a system.
One very common method to introduce dissipation into the real-time dynamics of quantum systems is to employ the Lindblad master equation~\cite{Manzano20} that extends the von Neumann equation by the additional term 
\begin{align}
    \mathcal{L}[\rho] = \sum_j \gamma_j(L_j \rho L_j^\dagger - \frac{1}{2}\{L_j^\dagger L_j, \rho\}),\label{lindgeneral}
\end{align}
where $\gamma_j$ are real-valued coefficients and $L_j$ are called Lindblad operators that capture the dissipative processes. 
While in principle the operators $L_j$ can be chosen freely, in quantum computing, quantum optical systems, spin-lattice dynamics, and other two-level ensembles, 
most commonly, local dissipative processes are considered and captured using the operators $\sigma_z$ and $\sigma_\pm=(\sigma_x \pm i\sigma_y)/2$.
For simplicity, we consider these processes to occur in every two-level system identically, i.e.\ with the same coefficients $\gamma_{z,\pm}$.
For the operator $\sigma_z$ in the ORQA formalism, the Lindblad action of the component corresponding to a multi-index $I$ is 
\begin{equation}
    {\mathcal{L}_{z}[\rho]}_I = -2 \gamma_z \rho_I \sum_{j=0}^{n-1} I_{j,1}\veebar I_{j,2},\label{lindZ}
\end{equation}
where $I_{j,1}$ and $I_{j,2}$ are the first and second bits of $I_j$, which is the $j$th two-bit string of $I$.
For $\sigma_\pm$ it is 
\begin{eqnarray}
    {\mathcal{L}_{\pm}[\rho]}_I &= \gamma_\pm(\pm \sum_{j=0}^{n-1}\delta_{I_j,3}\rho_{I \veebar 3 \times 4^j}-\frac{\oplus I}{2} \rho_I ),\label{lindPM}
\end{eqnarray}
where $\oplus$ denotes the binary digit sum, e.g.\ $\oplus 1011_2 = 3$.
Note that the actions of these Lindblad terms require very little access to other components of $\rho$, which makes it a very efficient subroutine.
We derive Eqs.~\ref{lindZ} and~\ref{lindPM} in App.~\ref{app-lindblad}.

\subsection{Imaginary-Time Evolution}\label{ITsection}

The imaginary-time evolution is a theoretical dynamical description that produces thermal states of systems in the density operator representation. 
The general idea relies on performing the transformation $\frac{i t}{\hbar}\rightarrow \beta$ with $\beta = {(k_B T)}^{-1}$, sometimes referred to as ``rotating time in the complex plane'', which intuitively transforms cyclical unitary dynamics to thermodynamic decay, with an exponential mapping on complex-valued objects.
The imaginary-time evolution therefore produces the normalized states
\begin{equation}
    \rho_T = \frac{\exp\{-\beta H\}}{\mathrm{Tr}(\exp\{-\beta H\})}.
\end{equation}
The von Neumann equation transforms in this case into the trace-preserving equation of motion
\begin{equation}
    \dot\rho = - \{\rho,H\} + 2 \mathrm{Tr}(H\rho)\rho.
\end{equation}
In the ORQA formalism, the expectation value of $H$ is 
\begin{equation}
    \mathrm{Tr}(H\rho) = {\sum_I}^\prime H_I\rho_I
\end{equation}
and the anti-commutator is calculated analogously to Eq.~\ref{repH} utilizing Eq.~\ref{anticom}, such that
\begin{align}
    \dot\rho_I &= 2{\sum_J}^\prime (\rho_{I\veebar J} X^{J, I\veebar J}+\rho_J)H_J\\
    &= 2{\sum_J}^\prime (H_{I\veebar J} X^{I\veebar J,J}+H_J)\rho_J.
\end{align}
Note that there are again two valid ways of writing this expression, depending on whether the sum is performed over the components of $\rho$ or those of $H$.
For long enough times, the relaxation dynamics provided by these equations approach the ground state of the Hamiltonian $H$, as the temperature approaches zero.

\subsection{Projective Measurements}\label{projsection}

A central part of quantum mechanics is the destructive measurement process which is captured with projections onto eigenstates.
A physical measurement~\footnote{Here we focus on measurements of the first kind, comp.~\cite{gardiner00}} with respect to some observable 
\begin{equation}
    O = \sum_i O_i P_i 
\end{equation}
produces one of the possible eigenvalues $O_i$ and projects the state into the corresponding eigenstate. 
$P_i$ are the projection operators of these corresponding eigenstates. 
The measurement process is described as transforming the density operator as given by 
\begin{equation}
    \rho \rightarrow \rho' = \frac{P_i \rho P_i}{\mathrm{Tr}(P_i\rho)},
\end{equation}
with probability $\mathrm{Tr}(P_i\rho)$.

It is also possible to consider performing such a measurement, but discarding the observed value, i.e.\ not ``rolling the dice``.
This provides a classically mixed state with the probabilities of observing each eigenvalue in the projected amplitudes.
In that case the density operator is transformed as 
\begin{equation}
    \rho \rightarrow \rho'=\sum_i P_i \rho P_i.
\end{equation}

In the particular case of the local observable $\sigma_k$ of the $j$th two-level system, the projectors onto the eigenstates of this operator are $P^\pm = (\sigma_0\pm\sigma_k)/2$ and produce the projected states
\begin{align}
    P_\pm \rho P_\pm 
    &=\frac{1}{2}{\sum_I}^\prime (\delta_{I_j,0}+\delta_{I_j,k})(\rho_I\pm\rho_{I\veebar k})\sigma_I,
\end{align}
where we have used that 
\begin{equation}
    \sigma_k\sigma_I\sigma_k=
    \begin{cases}
        +\sigma_I, &k=0\lor k=I_j\\
        -\sigma_I, &\mathrm{else}
    \end{cases}.
\end{equation}
A discarded measurement with respect to this operator therefore acts on the density operator as
\begin{align}
    \rho 
    &\rightarrow P^+ \rho P^+ + P^- \rho P^-\\
    &= {\sum_I}^\prime \rho_I \sigma_I (\delta_{I_j,0} + \delta_{I_j,k}).
\end{align}
The measurement probability of either state is given by 
\begin{equation}
    p_\pm = \frac{1}{2}\mathrm{Tr}(\rho \pm \sigma_k\rho) = \frac{1\pm\rho_k}{2}.
\end{equation}

Further, it is possible to consider the process of measuring parts of a quantum system, but discarding not only the information about the outcome, but also the information about the measurement operator.
This then puts the measured subsystem into a classical maximal entropy state which is mathematically captured by taking the partial trace over the measured subsystem.
We discuss the partial trace in the ORQA formalism in App.~\ref{app-partial}.

\subsection{Dynamic Truncation}\label{dyntran}

In order to dynamically reduce the number $|\rho|$ of components of the density operator, we introduce a straight-forward truncation method that limits the number of non-zero components of $\rho$. 
Since the number of components in $\rho$ determines the scaling of the implementation, we aim to discard components in $\rho$ which are close to zero, as they do not affect the dynamics significantly, but are numerically expensive. 
We do this by employing a truncation criterion.
After each iteration of updating $\rho$, e.g.\ via the Lindblad master equation, we remove all components of $\rho$ with absolute values smaller than a threshold $\epsilon>0$, i.e.\ 
\begin{equation}
    \rho_I \rightarrow \rho_I' = 
    \begin{cases}
        \rho_I, &|\rho_I|>\epsilon\\
        0, &|\rho_I|\leq\epsilon
    \end{cases}.
\end{equation}
Here setting $\rho_I$ to zero implies discarding the corresponding components from memory in the numerical implementation.
The result is an approximation of $\rho$ that is increasingly sparse with $\epsilon$.
We illustrate this truncation step in Fig.~\ref{cartoon}~(c).
Note, that this method of truncation is very straight-forward compared to other potential approaches.
In this method, the creation of new components $\rho_I$ relies on the details of the numerical integration and in particular the time-discretization $\Delta t$, as updates $|\dot\rho_I|\Delta t < \epsilon$ may effectively be rejected. 
This leads to a non-trivial interplay between $\Delta t$ and $\epsilon$.
However, this does not significantly affect the results in Section~3.
We consider the method by which the dynamic size of $\rho$ is controlled to be an important aspect of future extensions and utilization of the ORQA formalism.

\section{Numerical Demonstration}\label{numdemo}

In this section we present a numerical demonstration of the ORQA formalism applied to the examples of simulated and adiabatic quantum annealing for the maximum independent set (MIS) problem. 
Throughout this numerical demonstration, we work in a unit-less description as the energy- and time-scales are secondary to the performance of our method.

The MIS problem is an NP-hard combinatorics problem on undirected graphs. 
An undirected graph $G=(V,E)$ is defined as a set of vertices $V$ connected via edges $E$. 
Here we consider edges that connect all pairs of vertices $V_i$ and $V_j$ for which $|V_i-V_j|<1$.
A sub-graph $S \subset G$ is called an independent set, if no two vertices in $S$ are connected.
A MIS of $G$ is an independent set with the largest amount $|S|$ of vertices of all possible independent sets of $G$.
The MIS problem consists of identifying a MIS of a given graph $G$.

\begin{figure}[t] 
    \centering 
    \includegraphics[width=0.9\linewidth]{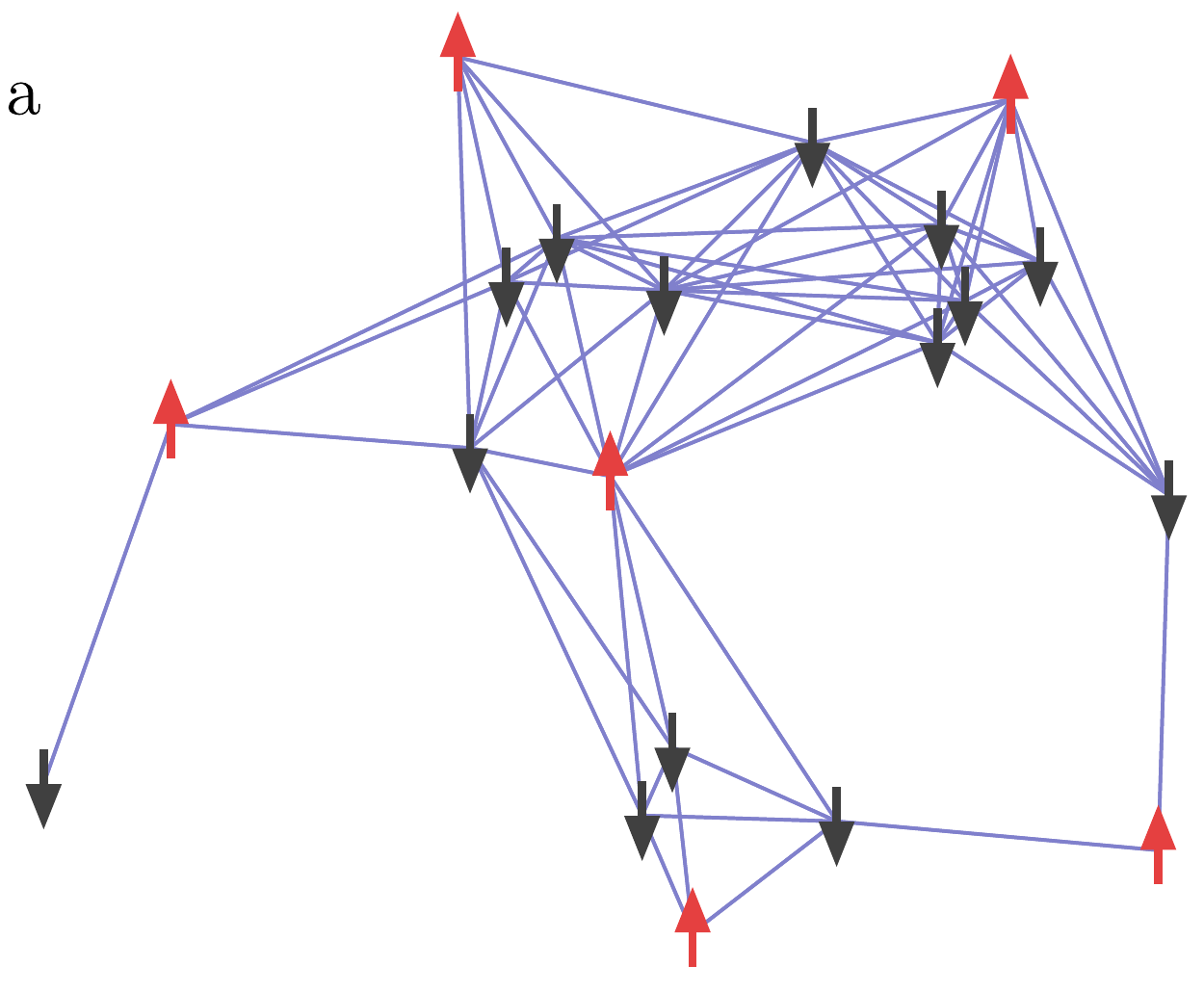}
    \includegraphics[width=1\linewidth]{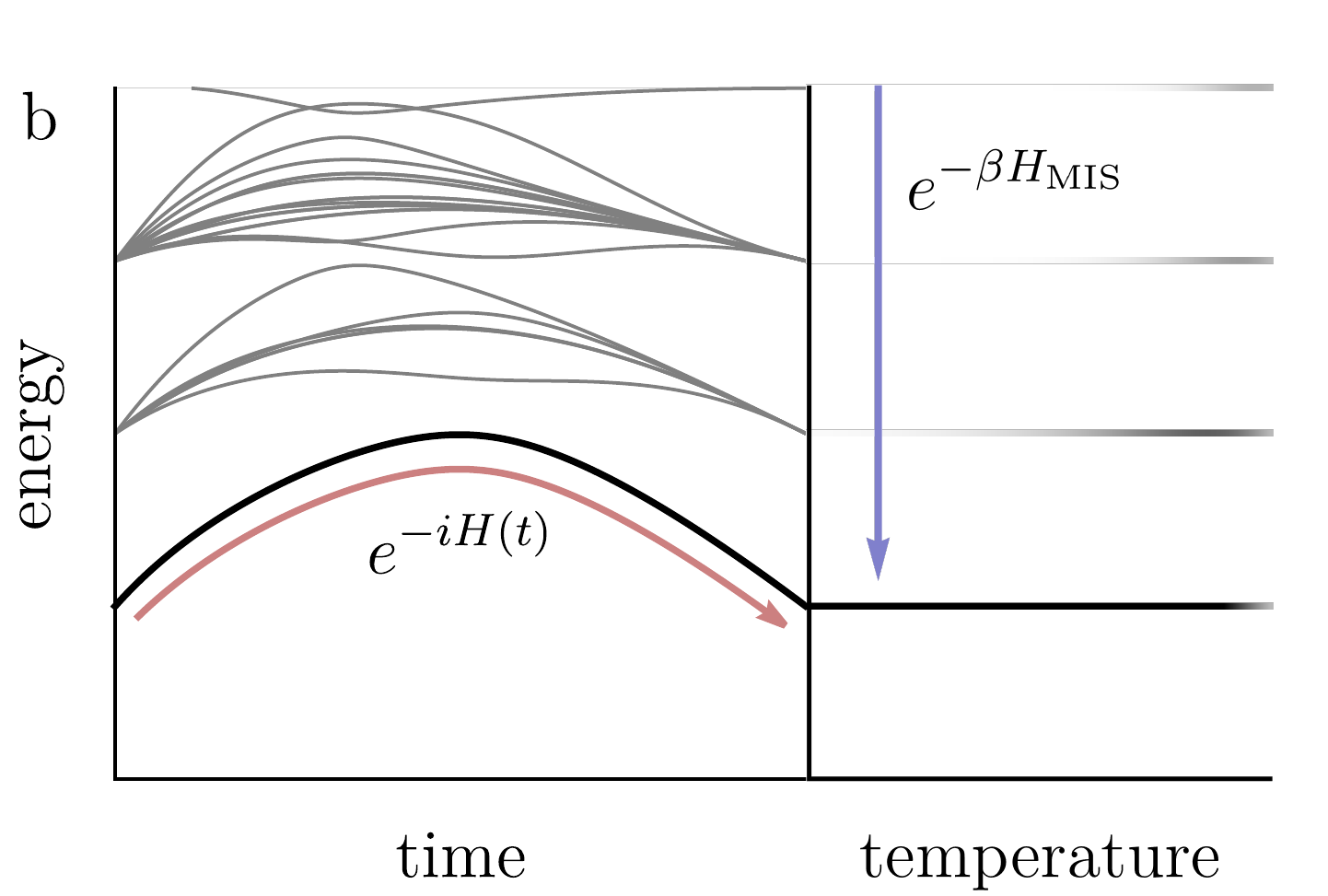}
    \caption{\textbf{Visualization of quantum annealing for the MIS problem.}
    Panel~(a) shows an example of an undirected graph with a MIS highlighted in red. 
    The set of all possible MISs is the space of degenerate ground states of the corresponding Hamiltonian $H_\mathrm{MIS}$ that encodes the given undirected graph.
    Panel~(b) illustrates the eigenvalues of the time-dependent Hamiltonian in Eq.~\ref{Hinterp} in adiabatic quantum annealing, where the slow interpolation in time $t$ from an initial Hamiltonian to the target Hamiltonian connects the ground states as indicated by the red arrow. 
    Panel~(c) illustrates the thermal states of $H_\mathrm{MIS}$ as a function of temperature $T$ in the imaginary-time evolution utilized in simulated quantum annealing.
    For vanishing temperatures the system approaches a classical mixture of the degenerate ground states as indicated by the blue arrow. 
    }\label{mis1}
\end{figure}

Like many problems in combinatorics, this can be mapped onto quadratic unconstrained binary optimization (QUBO)~\cite{glover19} and in particular onto Ising-type models~\cite{lucas14} in a way that the ground state of the resulting Hamiltonian encodes the solution to the problem, in this case the MIS\@.
This is a central use-case of quantum annealing and has been demonstrated in physical realizations of quantum annealing machines~\cite{willsch22,Ebadi22}.
We consider the mapping of the MIS problem onto the Hamiltonian
\begin{align}
    H_\mathrm{MIS} &= \sum_{i=1,j>i}^{|G|} P_\uparrow^i \otimes P_\uparrow^j \Theta_{ij} - \frac{1}{2}\sum_{i=1}^{|G|} P_\uparrow^i,\label{HMIS}
\end{align}
where $\Theta_{ij}=\Theta(|V_i-V_j|-1)$ is the Heaviside step-function and $P_\uparrow = \frac{1}{2}(\sigma_0+\sigma_z)$.
The degenerate ground state of this Hamiltonian is guaranteed to represent exactly all MISs of $G$, when the set of spins in the up-state in a given state represents the elements of a potentially independent set, see App.~\ref{app-groundstate}.
We show an example of an undirected graph and a MIS for twenty vertices in Fig.~\ref{mis1}~(a).

We prepare the ground state of this Hamiltonian via two different quantum annealing approaches implemented numerically in the ORQA formalism.
We choose this particular example as a proof of concept demonstration, because it is numerically hard and exact solutions scale exponentially with system size, while it also allows us to apply the imaginary-time evolution as well as the real-time evolution to the exact same problem. 
Note that the ORQA formalism is general and can very well be applied to systems with less connectivity and a smaller effective Hilbert-space, in which case the dynamics directly benefit from the scaling behavior of ORQA.

First, we demonstrate simulated quantum annealing through the imaginary-time evolution to infer the ground state as described in Section~\ref{ITsection}. 
We illustrate the process of imaginary-time evolution into the ground state in Fig.~\ref{mis1}~(c).
Second, we demonstrate adiabatic quantum annealing through the real-time evolution in the presence of dissipation as described in Sections~\ref{RTsection} and~\ref{lindsection}.
We illustrate the process of adiabatically transforming the ground state in Fig.~\ref{mis1}~(b).

After approaching the approximate ground state via each method, we proceed to systematically project the system in a measurement process as we described in Section~\ref{projsection}. 
We project the spin with the largest expectation value onto the up-state, and all $n_j$ connected spins into the down-state. 
We repeat this process for the reduced density operator of the subset of $G/R_j$ containing the remaining $n-n_j-1$ spins.
Here 
\begin{equation} 
    R_j=V_j\cup\{V_i \in G: |V_j-V_i|<1\}
\end{equation} 
is the set of $V_j$ and its neighbors.
After less than $n/2$ iterations of this projection process, all spins have been evaluated, and the system has been fully projected into a state composed of up- and down-spins that encode an independent set.
This state is guaranteed to be an MIS if we perform this projection procedure onto a mixture of the degenerate ground states. 
However, even in the case of an arbitrary state, this procedure produces an independent set, which may be an MIS by chance.
This method of identifying a MIS could therefore be modified to yield increased success rates by introducing means to redo the projection heuristically.
However, the focus of this section is not to introduce effective greedy heuristics, but to demonstrate the numerical performance of the ORQA formalism.
Note that the results we present here are performed without any parallelization of the numerical method, which would improve performance further. 
We leave this to future work. 
Here, we choose to implement the integration steps of ORQA in a $4$th-order Runke-Kutta method.

\subsection{Simulated Quantum Annealing}\label{exMIS}

\begin{figure*}[t]
    \centering  
    \includegraphics[width=\linewidth]{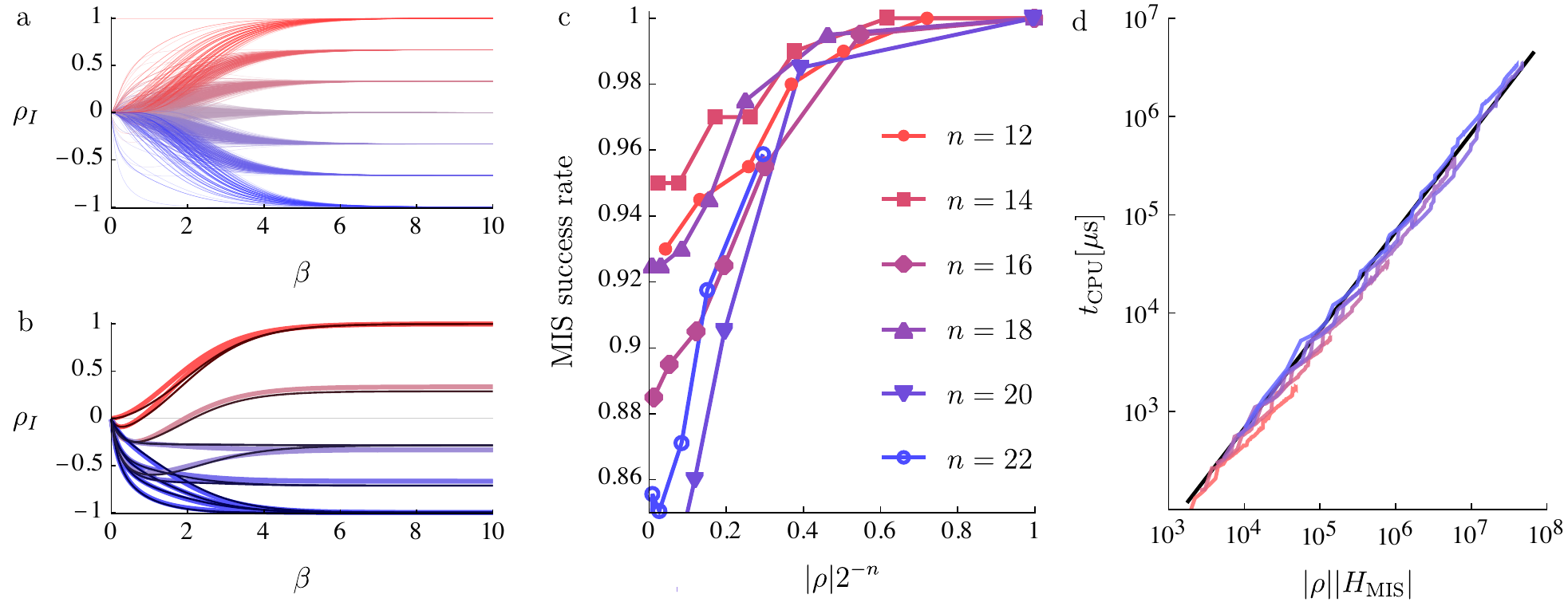}
    \caption{
        \textbf{Performance and scaling of simulated quantum annealing.}
        Panel~(a) shows the components $\rho_I$ of the thermal states for an example graph as a function of $\beta$ for $n=12$ and without dynamic truncation, i.e.\ $\epsilon=0$. 
        The color of each line is given by its asymptotic value for large $\beta$ (low temperatures), i.e.\ towards the ground state.
        Panel~(b) shows the same dynamics, but only for the $\rho_I$ corresponding to local single-body $\sigma_z$ terms.
        Additionally, the thin dark lines show the same $\rho_I$ in the presence of dynamic truncation with $\epsilon=0.0005$.
        Panel~(c) shows the success rate of identifying MISs, averaged over $200$ randomly generated graphs, in relation to the size of $\rho$.
        The individual lines correspond to system sizes up to $n=22$, indicated by colors, and individual dots correspond to different values of the truncation threshold up to $\epsilon=0.003$.
        Note that we do not show the case of $\epsilon=0$ for $n=22$.
        Panel~(d) shows the computational time $t_\mathrm{CPU}$ that is required to perform a single integration step as a function of the effective complexity $|\rho||H_\mathrm{MIS}|$ for $\epsilon=0.0015$.
        The lines correspond to individual trajectories with colors that correspond to the same system sizes as Panel (c).
        The black line acts as a visual guide.
        The dependence of the computational time $t_\mathrm{CPU}$ is approximately linear in the complexity $|\rho| |H_\mathrm{MIS}|$.
    }\label{misImagtime} 
\end{figure*}

As our first example, we consider simulated quantum annealing where we infer the ground state of $H_\mathrm{MIS}$ from vanishing-temperature states that we obtain through imaginary-time evolution.
We consider the initial infinite-temperature state 
\begin{equation}
    \rho(t=0)=\lim_{T\rightarrow\infty} \frac{e^{-\beta H_\mathrm{MIS}}}{\mathrm{Tr}(e^{-\beta H_\mathrm{MIS}})} = \mathbb{1}_{2^{n}}=\sigma_0
\end{equation}
with $\beta=(k_B T)^{-1}$,
and perform the imaginary-time evolution in the ORQA formalism for system sizes up to $n=22$ on randomly generated undirected graphs for varying values of $\epsilon$.
We analyze the success rate of identifying the MIS, and the effective system size $|\rho|$ under consideration of dynamic truncation as described in Section~\ref{dyntran}, and present the results in Fig.~\ref{misImagtime}.
We also show an example of the imaginary-time dynamics of $\rho$ with and without truncation.

Fig.~\ref{misImagtime}~(a) shows all components $\rho_I$ during the imaginary-time evolution for an example undirected graph in the absence of truncation, i.e.\ $\epsilon=0$.
During the transient, the $\rho_I$ spread out before they approach a discrete set of asymptotic values.
The color of each trajectory is given by the asymptotic value it approaches.
Note that these discrete values reflect the number of MIS solutions, as these expectation values are the ratio of MIS solutions that include the corresponding spins.
In principle, the ground state can be inferred as soon as the individual $\rho_I$ can be accurately assigned into clusters that approach the same value, which may happen early on in the time-evolution.
Fig.~\ref{misImagtime}~(b) displays the same dynamics, but shows only the subset of the $\rho_I$ that belong to local $\sigma_z$ expectation values, i.e.\ $I=3\times 4^j$.
The dark lines show the same trajectories in the presence of a non-zero truncation threshold $\epsilon=0.0005$.
We note that the approximate solutions agree well with the exact solutions. 
With increasing $\epsilon$ this accuracy is subsequently lost.

The truncation leads to a significant reduction in the number of components in $\rho$, which directly translates into a speed-up of the numerics. 
This demonstrates how the scaling behavior of the ORQA formalism in combination with the dynamic truncation can provide accurate results at significantly increased efficiency.
We quantify this in Fig.~\ref{misImagtime}~(c), where we show the success rate of identifying MISs with respect to the size of $\rho$ for varying system sizes $n$ and truncation thresholds up to $\epsilon=0.003$. 
A larger value of $\epsilon$ leads to less accuracy and a smaller size of $\rho$.
However, we demonstrate that the truncation can be used to considerably reduce the complexity of $\rho$ without significantly compromising the success rate of solving the MIS problem. 
As it is expected at some point, for increasing values of $\epsilon$ the solutions decrease in accuracy leading to errors in capturing the ground state and therefore failing to infer the MIS\@.
The asymptotic values of the success rate for very small sizes $|\rho|$ shows the accuracy of identifying MISs from first-order approximations of thermal states.

In Fig.~\ref{misImagtime}~(d), we show the computational time $t_\mathrm{CPU}$ that is required to calculate single integration steps as a function of the effective system complexity quantified as $|\rho||H_\mathrm{MIS}|$.
Note that the value of $|H_\mathrm{MIS}|$ depends on the given undirected graph.
The individual lines correspond to integration trajectories for system sizes up to $n=22$.
The corresponding colors are the same as in Fig.~\ref{misImagtime}~(c).
Note that the exact system sizes and truncation thresholds are not crucial here.
More important is the effective size $|\rho|$ that emerges due to the dynamic truncation.
We find on this log-log scale that the functional dependency of $t_\mathrm{CPU}$ appears to be very close to linear, which is in good agreement with our general estimate of the complexity as $\mathcal{O}(|\rho||H|)$.
These results correspond to calculation times of about $50\mathrm{ns}$ per component in $\rho$ and $H_\mathrm{MIS}$, which is indicated by the black line.
 
\subsection{Adiabatic Quantum Annealing}\label{realMIS}
\begin{figure*}[t]
    \centering 
    \includegraphics[width=\linewidth]{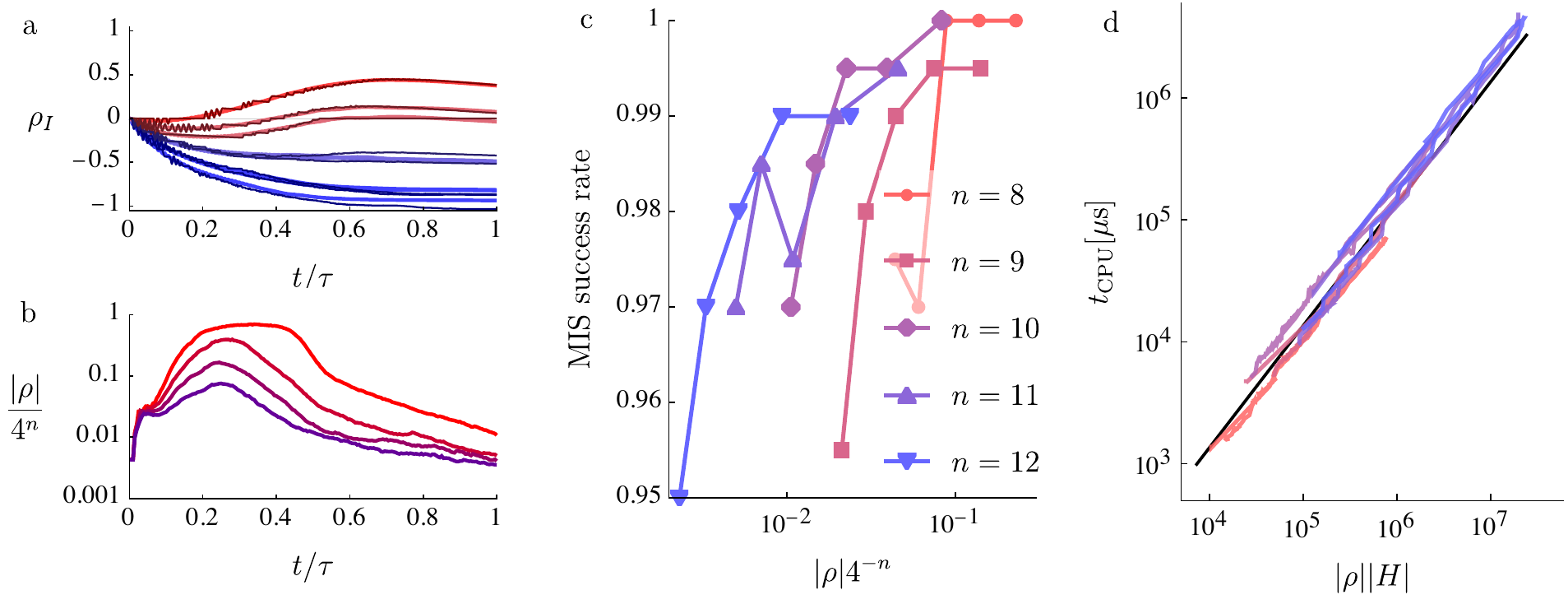}
    \caption{
        \textbf{Performance and scaling of dissipative adiabatic quantum annealing.}
        Panel~(a) shows the components $\rho_I$ corresponding to local $\sigma_z$ expectation values for a system of size $n=8$ for exact ($\epsilon=0$, colors) and truncated ($\epsilon=0.003$, thin dark) real-time evolution with $\gamma=0.25\tau^{-1}$.
        Panel~(b) shows the dynamically truncated size of $\rho$ during the real-time evolution for $\epsilon=0.003$ and dissipation with values from $\gamma=0.25\tau^{-1}$ to $\gamma=\tau^{-1}$.
        Panel~(c) shows the success rate of identifying MISs in relation to the size of $\rho$, averaged over $100$ to $200$ random graphs, depending on the system size which is indicated by color. 
        The individual lines correspond to different system sizes up to $n=12$ and truncation thresholds from $\epsilon=0.001$ up to $0.005$ for dissipation of $\gamma=0.5\tau^{-1}$.
        Panel~(d) shows the computational time $t_\mathrm{CPU}$ that is required to perform a single integration step as a function of the effective complexity $|\rho||H|$ for $\epsilon=0.003$ and $\gamma=0.5\tau^{-1}$.
        The lines correspond to trajectories with colors that correspond to the same system sizes as Panel (c).
        The black line acts as a visual guide. 
        The computational time $t_\mathrm{CPU}$ increases approximately linearly with $|\rho| |H|$.
    }\label{MISrealtime}
\end{figure*}

As our second example, we consider adiabatic quantum annealing, where we approach the ground state of the problem Hamiltonian $H_\mathrm{MIS}$ through dissipative real-time evolution. 
We initialize the system in the ground state of a simple Hamiltonian and then slowly transform this Hamiltonian into the problem Hamiltonian that encodes the MIS problem.
For this we write the time-dependent Hamiltonian 
\begin{equation}
    H(t) = H_\mathrm{MIS} \Gamma(t) + (1-\Gamma(t)) \sum_{j=0}^{n-1} \sigma_x^j,\label{Hinterp}
\end{equation}
with the initial ground state 
\begin{equation} 
    \rho(t=0) = \frac{1}{2^{n}}{(\sigma_0-\sigma_x)}^{\otimes n}.
\end{equation}
$\Gamma(t)$ is a function that slowly varies from $0$ to $1$. 
For simplicity, we choose $\Gamma(t)=\frac{t}{\tau}$, where $\tau$ is some final time.
In the case that at each point in time the gap to the first excited state does not close, then by the adiabatic theorem the final state $\rho(t=\tau)$ of the system will be the ground state of the problem Hamiltonian.

In terms of numerical resources, the real-time evolution of this Hamiltonian is more demanding than the imaginary-time evolution of $H_\mathrm{MIS}$, as it approaches the full complexity of $|\rho|=4^n$ and requires adiabatic dynamics.
Therefore, we consider this system for up to the size $n=12$ in the presence of dissipation by fully simulating the Lindblad master equation.
For simplicity, we consider a single dissipation coefficient $\gamma=\gamma_z=\gamma_-$ and $\gamma_+=0$. 
Analogously to Fig.~\ref{misImagtime} and the analysis of the previous subsection, in Fig.~\ref{MISrealtime} we show the performance of adiabatic quantum annealing in the presence of dissipation. 

In Fig.~\ref{MISrealtime}~(a), we show the local expectation values $\rho_{3\times 4^j} = \langle \sigma_z^j \rangle$ as functions of time for an example graph of size $n=8$. 
We show these expectation values in the presence of dissipation with $\gamma=0.25\tau^{-1}$.
The colors correspond to the values of the expectation values for $t=\tau$ in the equivalent dynamics in the absence of dissipation.
Analogously to the results of simulated quantum annealing in Section~\ref{exMIS}, the components $\rho_I$ in the final state approach a set of discrete values.
However, the values of the final state in the presence of dissipation are suppressed by exponential decay, which obscures valuable information about the ground state of $H_\mathrm{MIS}$.
Note, that the expectation values of the ground state can still be inferred in the presence of dissipation to a certain extent.
The dark lines show the same dynamics but in the presence of dynamic truncation with the threshold $\epsilon=0.003$.
While the truncated dynamics approximate the exact results fairly well for this particular set of parameters, we note that the adiabatic dynamics are susceptible to the truncation as can be seen in the minor oscillations emerging in the approximate solutions.

In Fig.~\ref{MISrealtime}~(b), we show the dynamic size of $\rho$, relative to the maximal value $4^n$, during the adiabatic quantum annealing process in the presence of truncation with the same value of $\epsilon=0.003$ as above.
We show $|\rho|$ for varying values of $\gamma$ and find that it reduces drastically due to the dissipation, as the components $\rho_I$ that are small and decay below the threshold $\epsilon$ are truncated.
Consequently, there is a general trend of increased complexity in $\rho$ emerging in the initial transient which then decays over time, as seen in the rise and fall of $|\rho|$ during the time-evolution.
We find that moderate dissipation leads to orders of magnitudes of reduced complexity in $\rho$, which directly leads to a speed-up of the approximate numerics with non-zero truncation thresholds.

In Fig.~\ref{MISrealtime}~(c), we show the success rate of identifying MISs after truncated dissipative adiabatic quantum annealing in relation to the number of components in $\rho$ for system sizes up to $n=12$.
Each line corresponds to a given system size, where each point corresponds to a different value of the truncation threshold from $\epsilon=0.001$ to $\epsilon=0.005$.
Here the dissipation coefficient is $\gamma=0.5\tau^{-1}$.
With increasing $\epsilon$, the size $|\rho|$ reduces quickly, which affects the accuracy of the calculation and at some point reduces the success rate.
In this sense, the dissipative real-time evolution is more sensitive to truncation, as the effective size is reduced by orders of magnitude.
Despite this drastic reduction, we find that the success rate remains high.
Note that randomly generated graphs display varying energy scales, such that for larger system sizes, some trajectories are potentially not fully adiabatic.
Further, the success rates corresponding to small values of $\epsilon$ are limited due to dissipation.

In Fig.~\ref{MISrealtime}~(d), we show the computational time $t_\mathrm{CPU}$ of single integration steps for dissipative adiabatic quantum annealing.
We show $t_\mathrm{CPU}$ as a function of the effective problem size $|\rho||H|$ on a log-log scale, and again find a close-to-linear dependence.
This is largely consistent with the results we show in Fig.~\ref{misImagtime}~(d).
Here $|H|=|H_\mathrm{MIS}|+n$, as given by Eq.~\ref{Hinterp}.
These results correspond to a computational time of about $100\mathrm{ns}$ per component in $\rho$ and $H$, which is depicted by the black line.  
Here, there appears a weak dependence on the system size $n$.
This largely depends on the exact numerical implementation of the objects $\rho$ and $H$, which contains overhead in obtaining the components $\rho_I$ and $H_I$ from memory. 
We consider the optimization of this aspect of our implementation to be a major opportunity for further optimization.

\section{Conclusion}\label{conclusion}

We have put forth a representation-free formalism of the algebraic structure of the Lie algebra $\mathfrak{su}(2^n)$.
We refer to this formalism as the exclusive-or represented quantum algebra (ORQA), as it reduces the algebraic structure to bit-wise exclusive-or operations on the enumeration-indices of elements of $\mathfrak{su}(2^n)$.
This method is general and has particular merit in utilization in the context of quantum many-body systems that are described as composites of two-level systems, used in quantum computation, quantum information theory at large, quantum optics, and spin-lattice models, to name a few.
While the analytical structure provided by ORQA is intriguing and at times surprisingly rich by itself, the benefits are particularly striking in numerical implementations of quantum dynamics, e.g.\ with the Lindblad-von Neumann master equation.
We have provided a proof of concept numerical demonstration of simulating up to 22 two-level systems in the example of quantum annealing in the combinatorics problem of identifying maximum independent sets in undirected graphs.
 
We find that the numerical implementation approximately scales as $\mathcal{O}(|\rho||H|)$, where $|\rho|$ and $|H|$ are the numbers of components used to represent the density operator and the Hamiltonian, respectively.
In most physical systems the Hamiltonian only contains a number of terms that is polynomial in the system size, such that the complexity is dominated by the linear dependence on the number of components in the density operator.
Naturally, in sufficiently complicated problems, this number will approach its upper bound of $4^n$, recovering the exponential complexity of quantum systems.
We find that a straight-forward truncation method that discards components of the density operator that are below a certain threshold in magnitude, already provides considerable speed-up while producing very good approximations of the exact solutions.
This advantage is particularly apparent in the Lindblad master equation, where dissipation can increase the performance by orders of magnitude.
 
The bottleneck of our implementation is the particular map that encodes the density operator and is read and written very frequently.
While numerical hash maps have a potentially constant-time access operation, this effectively increases in very large maps depending on the details of the implementation. 
We believe that a designated or problem-specific map can considerably boost the performance of our implementation further.
Additionally, the numerical implementation of the equations of motion such as the Lindblad master equation display a very large degree of parallelizability in the ORQA formalism.
We have not made use of that here, but parallelization promises to drastically increase the performance in future work. 
Unsurprisingly, memory requirements are a dominant obstacle, which is why we consider extensions of our dynamic truncation method to be an equally relevant aspect of future endeavors.

In summary, ORQA presents a completely general framework in the context of composites of two-level quantum systems, which are found all throughout quantum many-body physics.
Hence, it can readily be utilized in connection with well-established methods of numerical quantum many-body physics, such as density matrix renormalization group techniques, tensor network approaches, or quantum Monte-Carlo methods.
Further, selected mean-field or truncated Wigner approaches may benefit from ORQA as well.
As such, it can be applied to a very large class of timely relevant problems. 
The binary nature of ORQA displays a conceptual connection to quadratic unconstrained binary optimization (QUBO) which is intimately linked to Ising models and quantum annealing.
We propose the ORQA formalism as a powerful addition to the tool-set of numerical quantum simulations. 
Extending this formalism to efficiently harness parallelization and optimized data-structures provides many opportunities for future work.

\begin{acknowledgments}
This work is funded by the Deutsche Forschungsgemeinschaft (DFG, German Research Foundation) -- SFB-925 -- project 170620586,
and the Cluster of Excellence `Advanced Imaging of Matter' (EXC 2056), Project No. 390715994.
The project is co-financed by ERDF of the European Union and by 'Fonds of the Hamburg Ministry of Science, Research, Equalities and Districts (BWFGB)’.
\end{acknowledgments} 

\section*{Code Availability}
Our numerical implementation of the ORQA formalism will be made available as a light-weight \texttt{C++} library in the weeks following the publication of this pre-print.

\bibliography{lit}

\begin{thebibliography}{40}%
\makeatletter
\providecommand \@ifxundefined [1]{%
 \@ifx{#1\undefined}
}%
\providecommand \@ifnum [1]{%
 \ifnum #1\expandafter \@firstoftwo
 \else \expandafter \@secondoftwo
 \fi
}%
\providecommand \@ifx [1]{%
 \ifx #1\expandafter \@firstoftwo
 \else \expandafter \@secondoftwo
 \fi
}%
\providecommand \natexlab [1]{#1}%
\providecommand \enquote  [1]{``#1''}%
\providecommand \bibnamefont  [1]{#1}%
\providecommand \bibfnamefont [1]{#1}%
\providecommand \citenamefont [1]{#1}%
\providecommand \href@noop [0]{\@secondoftwo}%
\providecommand \href [0]{\begingroup \@sanitize@url \@href}%
\providecommand \@href[1]{\@@startlink{#1}\@@href}%
\providecommand \@@href[1]{\endgroup#1\@@endlink}%
\providecommand \@sanitize@url [0]{\catcode `\\12\catcode `\$12\catcode
  `\&12\catcode `\#12\catcode `\^12\catcode `\_12\catcode `\%12\relax}%
\providecommand \@@startlink[1]{}%
\providecommand \@@endlink[0]{}%
\providecommand \url  [0]{\begingroup\@sanitize@url \@url }%
\providecommand \@url [1]{\endgroup\@href {#1}{\urlprefix }}%
\providecommand \urlprefix  [0]{URL }%
\providecommand \Eprint [0]{\href }%
\providecommand \doibase [0]{https://doi.org/}%
\providecommand \selectlanguage [0]{\@gobble}%
\providecommand \bibinfo  [0]{\@secondoftwo}%
\providecommand \bibfield  [0]{\@secondoftwo}%
\providecommand \translation [1]{[#1]}%
\providecommand \BibitemOpen [0]{}%
\providecommand \bibitemStop [0]{}%
\providecommand \bibitemNoStop [0]{.\EOS\space}%
\providecommand \EOS [0]{\spacefactor3000\relax}%
\providecommand \BibitemShut  [1]{\csname bibitem#1\endcsname}%
\let\auto@bib@innerbib\@empty
\bibitem [{\citenamefont {Watrous}(2018)}]{Watrous18}%
  \BibitemOpen
  \bibfield  {author} {\bibinfo {author} {\bibfnamefont {J.}~\bibnamefont
  {Watrous}},\ }\href@noop {} {\emph {\bibinfo {title} {The Theory of Quantum
  Information}}}\ (\bibinfo  {publisher} {Cambridge University Press},\
  \bibinfo {year} {2018})\BibitemShut {NoStop}%
\bibitem [{\citenamefont {Nielsen}\ and\ \citenamefont
  {Chuang}(2010)}]{NielsenChuang10}%
  \BibitemOpen
  \bibfield  {author} {\bibinfo {author} {\bibfnamefont {M.~A.}\ \bibnamefont
  {Nielsen}}\ and\ \bibinfo {author} {\bibfnamefont {I.~L.}\ \bibnamefont
  {Chuang}},\ }\href@noop {} {\emph {\bibinfo {title} {Quantum Computation and
  Quantum Information: 10th Anniversary Edition}}}\ (\bibinfo  {publisher}
  {Cambridge University Press},\ \bibinfo {year} {2010})\BibitemShut {NoStop}%
\bibitem [{\citenamefont {Gross}\ and\ \citenamefont
  {Haroche}(1982)}]{gross82}%
  \BibitemOpen
  \bibfield  {author} {\bibinfo {author} {\bibfnamefont {M.}~\bibnamefont
  {Gross}}\ and\ \bibinfo {author} {\bibfnamefont {S.}~\bibnamefont
  {Haroche}},\ }\bibfield  {title} {\bibinfo {title} {Superradiance: An essay
  on the theory of collective spontaneous emission},\ }\href
  {https://doi.org/https://doi.org/10.1016/0370-1573(82)90102-8} {\bibfield
  {journal} {\bibinfo  {journal} {Physics Reports}\ }\textbf {\bibinfo {volume}
  {93}},\ \bibinfo {pages} {301} (\bibinfo {year} {1982})}\BibitemShut
  {NoStop}%
\bibitem [{\citenamefont {Temnov}\ and\ \citenamefont
  {Woggon}(2005)}]{temnov05}%
  \BibitemOpen
  \bibfield  {author} {\bibinfo {author} {\bibfnamefont {V.~V.}\ \bibnamefont
  {Temnov}}\ and\ \bibinfo {author} {\bibfnamefont {U.}~\bibnamefont
  {Woggon}},\ }\bibfield  {title} {\bibinfo {title} {Superradiance and
  subradiance in an inhomogeneously broadened ensemble of two-level systems
  coupled to a low-$q$ cavity},\ }\href
  {https://doi.org/10.1103/PhysRevLett.95.243602} {\bibfield  {journal}
  {\bibinfo  {journal} {Phys. Rev. Lett.}\ }\textbf {\bibinfo {volume} {95}},\
  \bibinfo {pages} {243602} (\bibinfo {year} {2005})}\BibitemShut {NoStop}%
\bibitem [{\citenamefont {Garraway}(2011)}]{garraway11}%
  \BibitemOpen
  \bibfield  {author} {\bibinfo {author} {\bibfnamefont {B.~M.}\ \bibnamefont
  {Garraway}},\ }\bibfield  {title} {\bibinfo {title} {The dicke model in
  quantum optics: Dicke model revisited},\ }\href
  {http://www.jstor.org/stable/41148867} {\bibfield  {journal} {\bibinfo
  {journal} {Philosophical Transactions: Mathematical, Physical and Engineering
  Sciences}\ }\textbf {\bibinfo {volume} {369}},\ \bibinfo {pages} {1137}
  (\bibinfo {year} {2011})}\BibitemShut {NoStop}%
\bibitem [{\citenamefont {Browne}\ \emph {et~al.}(2017)\citenamefont {Browne},
  \citenamefont {Bose}, \citenamefont {Mintert},\ and\ \citenamefont
  {Kim}}]{Browne17}%
  \BibitemOpen
  \bibfield  {author} {\bibinfo {author} {\bibfnamefont {D.}~\bibnamefont
  {Browne}}, \bibinfo {author} {\bibfnamefont {S.}~\bibnamefont {Bose}},
  \bibinfo {author} {\bibfnamefont {F.}~\bibnamefont {Mintert}},\ and\ \bibinfo
  {author} {\bibfnamefont {M.}~\bibnamefont {Kim}},\ }\bibfield  {title}
  {\bibinfo {title} {From quantum optics to quantum technologies},\ }\href
  {https://doi.org/https://doi.org/10.1016/j.pquantelec.2017.06.002} {\bibfield
   {journal} {\bibinfo  {journal} {Progress in Quantum Electronics}\ }\textbf
  {\bibinfo {volume} {54}},\ \bibinfo {pages} {2} (\bibinfo {year} {2017})},\
  \bibinfo {note} {special issue in honor of the 70th birthday of Professor Sir
  Peter Knight FRS}\BibitemShut {NoStop}%
\bibitem [{\citenamefont {Affleck}(1989)}]{Affleck89}%
  \BibitemOpen
  \bibfield  {author} {\bibinfo {author} {\bibfnamefont {I.}~\bibnamefont
  {Affleck}},\ }\bibfield  {title} {\bibinfo {title} {Quantum spin chains and
  the haldane gap},\ }\href {https://doi.org/10.1088/0953-8984/1/19/001}
  {\bibfield  {journal} {\bibinfo  {journal} {Journal of Physics: Condensed
  Matter}\ }\textbf {\bibinfo {volume} {1}},\ \bibinfo {pages} {3047} (\bibinfo
  {year} {1989})}\BibitemShut {NoStop}%
\bibitem [{\citenamefont {Khaneja}\ and\ \citenamefont
  {Glaser}(2001)}]{Khaneja01}%
  \BibitemOpen
  \bibfield  {author} {\bibinfo {author} {\bibfnamefont {N.}~\bibnamefont
  {Khaneja}}\ and\ \bibinfo {author} {\bibfnamefont {S.~J.}\ \bibnamefont
  {Glaser}},\ }\bibfield  {title} {\bibinfo {title} {Cartan decomposition of
  su(2n) and control of spin systems},\ }\href
  {https://doi.org/https://doi.org/10.1016/S0301-0104(01)00318-4} {\bibfield
  {journal} {\bibinfo  {journal} {Chemical Physics}\ }\textbf {\bibinfo
  {volume} {267}},\ \bibinfo {pages} {11} (\bibinfo {year} {2001})}\BibitemShut
  {NoStop}%
\bibitem [{\citenamefont {Backens}\ \emph {et~al.}(2019)\citenamefont
  {Backens}, \citenamefont {Shnirman},\ and\ \citenamefont
  {Makhlin}}]{Backens19}%
  \BibitemOpen
  \bibfield  {author} {\bibinfo {author} {\bibfnamefont {S.}~\bibnamefont
  {Backens}}, \bibinfo {author} {\bibfnamefont {A.}~\bibnamefont {Shnirman}},\
  and\ \bibinfo {author} {\bibfnamefont {Y.}~\bibnamefont {Makhlin}},\
  }\bibfield  {title} {\bibinfo {title} {Jordan--wigner transformations for
  tree structures},\ }\href {https://doi.org/10.1038/s41598-018-38128-8}
  {\bibfield  {journal} {\bibinfo  {journal} {Scientific Reports}\ }\textbf
  {\bibinfo {volume} {9}},\ \bibinfo {pages} {2598} (\bibinfo {year}
  {2019})}\BibitemShut {NoStop}%
\bibitem [{\citenamefont {Jordan}\ and\ \citenamefont
  {Wigner}(1928)}]{Jordan28}%
  \BibitemOpen
  \bibfield  {author} {\bibinfo {author} {\bibfnamefont {P.}~\bibnamefont
  {Jordan}}\ and\ \bibinfo {author} {\bibfnamefont {E.}~\bibnamefont
  {Wigner}},\ }\bibfield  {title} {\bibinfo {title} {{\"U}ber das paulische
  {\"a}quivalenzverbot},\ }\href {https://doi.org/10.1007/BF01331938}
  {\bibfield  {journal} {\bibinfo  {journal} {Zeitschrift f{\"u}r Physik}\
  }\textbf {\bibinfo {volume} {47}},\ \bibinfo {pages} {631} (\bibinfo {year}
  {1928})}\BibitemShut {NoStop}%
\bibitem [{\citenamefont {Schwinger}(2015)}]{schwinger52}%
  \BibitemOpen
  \bibfield  {author} {\bibinfo {author} {\bibfnamefont {J.}~\bibnamefont
  {Schwinger}},\ }\href {https://books.google.de/books?id=LF80BwAAQBAJ} {\emph
  {\bibinfo {title} {On Angular Momentum}}},\ Dover Books on Physics\ (\bibinfo
   {publisher} {Dover Publications},\ \bibinfo {year} {2015})\BibitemShut
  {NoStop}%
\bibitem [{\citenamefont {Holstein}\ and\ \citenamefont
  {Primakoff}(1940)}]{Holstein40}%
  \BibitemOpen
  \bibfield  {author} {\bibinfo {author} {\bibfnamefont {T.}~\bibnamefont
  {Holstein}}\ and\ \bibinfo {author} {\bibfnamefont {H.}~\bibnamefont
  {Primakoff}},\ }\bibfield  {title} {\bibinfo {title} {Field dependence of the
  intrinsic domain magnetization of a ferromagnet},\ }\href
  {https://doi.org/10.1103/PhysRev.58.1098} {\bibfield  {journal} {\bibinfo
  {journal} {Phys. Rev.}\ }\textbf {\bibinfo {volume} {58}},\ \bibinfo {pages}
  {1098} (\bibinfo {year} {1940})}\BibitemShut {NoStop}%
\bibitem [{\citenamefont {Georgescu}\ \emph {et~al.}(2014)\citenamefont
  {Georgescu}, \citenamefont {Ashhab},\ and\ \citenamefont
  {Nori}}]{Georgescu14}%
  \BibitemOpen
  \bibfield  {author} {\bibinfo {author} {\bibfnamefont {I.~M.}\ \bibnamefont
  {Georgescu}}, \bibinfo {author} {\bibfnamefont {S.}~\bibnamefont {Ashhab}},\
  and\ \bibinfo {author} {\bibfnamefont {F.}~\bibnamefont {Nori}},\ }\bibfield
  {title} {\bibinfo {title} {Quantum simulation},\ }\href
  {https://doi.org/10.1103/RevModPhys.86.153} {\bibfield  {journal} {\bibinfo
  {journal} {Rev. Mod. Phys.}\ }\textbf {\bibinfo {volume} {86}},\ \bibinfo
  {pages} {153} (\bibinfo {year} {2014})}\BibitemShut {NoStop}%
\bibitem [{\citenamefont {Altman}\ \emph {et~al.}(2021)\citenamefont {Altman},
  \citenamefont {Brown}, \citenamefont {Carleo}, \citenamefont {Carr},
  \citenamefont {Demler}, \citenamefont {Chin}, \citenamefont {DeMarco},
  \citenamefont {Economou}, \citenamefont {Eriksson}, \citenamefont {Fu},
  \citenamefont {Greiner}, \citenamefont {Hazzard}, \citenamefont {Hulet},
  \citenamefont {Koll\'ar}, \citenamefont {Lev}, \citenamefont {Lukin},
  \citenamefont {Ma}, \citenamefont {Mi}, \citenamefont {Misra}, \citenamefont
  {Monroe}, \citenamefont {Murch}, \citenamefont {Nazario}, \citenamefont {Ni},
  \citenamefont {Potter}, \citenamefont {Roushan}, \citenamefont {Saffman},
  \citenamefont {Schleier-Smith}, \citenamefont {Siddiqi}, \citenamefont
  {Simmonds}, \citenamefont {Singh}, \citenamefont {Spielman}, \citenamefont
  {Temme}, \citenamefont {Weiss}, \citenamefont {Vu\ifmmode \check{c}\else
  \v{c}\fi{}kovi\ifmmode~\acute{c}\else \'{c}\fi{}}, \citenamefont
  {Vuleti\ifmmode~\acute{c}\else \'{c}\fi{}}, \citenamefont {Ye},\ and\
  \citenamefont {Zwierlein}}]{Altman21}%
  \BibitemOpen
  \bibfield  {author} {\bibinfo {author} {\bibfnamefont {E.}~\bibnamefont
  {Altman}}, \bibinfo {author} {\bibfnamefont {K.~R.}\ \bibnamefont {Brown}},
  \bibinfo {author} {\bibfnamefont {G.}~\bibnamefont {Carleo}}, \bibinfo
  {author} {\bibfnamefont {L.~D.}\ \bibnamefont {Carr}}, \bibinfo {author}
  {\bibfnamefont {E.}~\bibnamefont {Demler}}, \bibinfo {author} {\bibfnamefont
  {C.}~\bibnamefont {Chin}}, \bibinfo {author} {\bibfnamefont {B.}~\bibnamefont
  {DeMarco}}, \bibinfo {author} {\bibfnamefont {S.~E.}\ \bibnamefont
  {Economou}}, \bibinfo {author} {\bibfnamefont {M.~A.}\ \bibnamefont
  {Eriksson}}, \bibinfo {author} {\bibfnamefont {K.-M.~C.}\ \bibnamefont {Fu}},
  \bibinfo {author} {\bibfnamefont {M.}~\bibnamefont {Greiner}}, \bibinfo
  {author} {\bibfnamefont {K.~R.}\ \bibnamefont {Hazzard}}, \bibinfo {author}
  {\bibfnamefont {R.~G.}\ \bibnamefont {Hulet}}, \bibinfo {author}
  {\bibfnamefont {A.~J.}\ \bibnamefont {Koll\'ar}}, \bibinfo {author}
  {\bibfnamefont {B.~L.}\ \bibnamefont {Lev}}, \bibinfo {author} {\bibfnamefont
  {M.~D.}\ \bibnamefont {Lukin}}, \bibinfo {author} {\bibfnamefont
  {R.}~\bibnamefont {Ma}}, \bibinfo {author} {\bibfnamefont {X.}~\bibnamefont
  {Mi}}, \bibinfo {author} {\bibfnamefont {S.}~\bibnamefont {Misra}}, \bibinfo
  {author} {\bibfnamefont {C.}~\bibnamefont {Monroe}}, \bibinfo {author}
  {\bibfnamefont {K.}~\bibnamefont {Murch}}, \bibinfo {author} {\bibfnamefont
  {Z.}~\bibnamefont {Nazario}}, \bibinfo {author} {\bibfnamefont {K.-K.}\
  \bibnamefont {Ni}}, \bibinfo {author} {\bibfnamefont {A.~C.}\ \bibnamefont
  {Potter}}, \bibinfo {author} {\bibfnamefont {P.}~\bibnamefont {Roushan}},
  \bibinfo {author} {\bibfnamefont {M.}~\bibnamefont {Saffman}}, \bibinfo
  {author} {\bibfnamefont {M.}~\bibnamefont {Schleier-Smith}}, \bibinfo
  {author} {\bibfnamefont {I.}~\bibnamefont {Siddiqi}}, \bibinfo {author}
  {\bibfnamefont {R.}~\bibnamefont {Simmonds}}, \bibinfo {author}
  {\bibfnamefont {M.}~\bibnamefont {Singh}}, \bibinfo {author} {\bibfnamefont
  {I.}~\bibnamefont {Spielman}}, \bibinfo {author} {\bibfnamefont
  {K.}~\bibnamefont {Temme}}, \bibinfo {author} {\bibfnamefont {D.~S.}\
  \bibnamefont {Weiss}}, \bibinfo {author} {\bibfnamefont {J.}~\bibnamefont
  {Vu\ifmmode \check{c}\else \v{c}\fi{}kovi\ifmmode~\acute{c}\else
  \'{c}\fi{}}}, \bibinfo {author} {\bibfnamefont {V.}~\bibnamefont
  {Vuleti\ifmmode~\acute{c}\else \'{c}\fi{}}}, \bibinfo {author} {\bibfnamefont
  {J.}~\bibnamefont {Ye}},\ and\ \bibinfo {author} {\bibfnamefont
  {M.}~\bibnamefont {Zwierlein}},\ }\bibfield  {title} {\bibinfo {title}
  {Quantum simulators: Architectures and opportunities},\ }\href
  {https://doi.org/10.1103/PRXQuantum.2.017003} {\bibfield  {journal} {\bibinfo
   {journal} {PRX Quantum}\ }\textbf {\bibinfo {volume} {2}},\ \bibinfo {pages}
  {017003} (\bibinfo {year} {2021})}\BibitemShut {NoStop}%
\bibitem [{\citenamefont {Finnila}\ \emph {et~al.}(1994)\citenamefont
  {Finnila}, \citenamefont {Gomez}, \citenamefont {Sebenik}, \citenamefont
  {Stenson},\ and\ \citenamefont {Doll}}]{Finnila94}%
  \BibitemOpen
  \bibfield  {author} {\bibinfo {author} {\bibfnamefont {A.}~\bibnamefont
  {Finnila}}, \bibinfo {author} {\bibfnamefont {M.}~\bibnamefont {Gomez}},
  \bibinfo {author} {\bibfnamefont {C.}~\bibnamefont {Sebenik}}, \bibinfo
  {author} {\bibfnamefont {C.}~\bibnamefont {Stenson}},\ and\ \bibinfo {author}
  {\bibfnamefont {J.}~\bibnamefont {Doll}},\ }\bibfield  {title} {\bibinfo
  {title} {Quantum annealing: A new method for minimizing multidimensional
  functions},\ }\href
  {https://doi.org/https://doi.org/10.1016/0009-2614(94)00117-0} {\bibfield
  {journal} {\bibinfo  {journal} {Chemical Physics Letters}\ }\textbf {\bibinfo
  {volume} {219}},\ \bibinfo {pages} {343} (\bibinfo {year}
  {1994})}\BibitemShut {NoStop}%
\bibitem [{\citenamefont {Hauke}\ \emph {et~al.}(2020)\citenamefont {Hauke},
  \citenamefont {Katzgraber}, \citenamefont {Lechner}, \citenamefont
  {Nishimori},\ and\ \citenamefont {Oliver}}]{Hauke20}%
  \BibitemOpen
  \bibfield  {author} {\bibinfo {author} {\bibfnamefont {P.}~\bibnamefont
  {Hauke}}, \bibinfo {author} {\bibfnamefont {H.~G.}\ \bibnamefont
  {Katzgraber}}, \bibinfo {author} {\bibfnamefont {W.}~\bibnamefont {Lechner}},
  \bibinfo {author} {\bibfnamefont {H.}~\bibnamefont {Nishimori}},\ and\
  \bibinfo {author} {\bibfnamefont {W.~D.}\ \bibnamefont {Oliver}},\ }\bibfield
   {title} {\bibinfo {title} {Perspectives of quantum annealing: methods and
  implementations},\ }\href {https://doi.org/10.1088/1361-6633/ab85b8}
  {\bibfield  {journal} {\bibinfo  {journal} {Reports on Progress in Physics}\
  }\textbf {\bibinfo {volume} {83}},\ \bibinfo {pages} {054401} (\bibinfo
  {year} {2020})}\BibitemShut {NoStop}%
\bibitem [{\citenamefont {Biamonte}\ and\ \citenamefont
  {Love}(2008)}]{Biamonte08}%
  \BibitemOpen
  \bibfield  {author} {\bibinfo {author} {\bibfnamefont {J.~D.}\ \bibnamefont
  {Biamonte}}\ and\ \bibinfo {author} {\bibfnamefont {P.~J.}\ \bibnamefont
  {Love}},\ }\bibfield  {title} {\bibinfo {title} {Realizable hamiltonians for
  universal adiabatic quantum computers},\ }\href
  {https://doi.org/10.1103/PhysRevA.78.012352} {\bibfield  {journal} {\bibinfo
  {journal} {Phys. Rev. A}\ }\textbf {\bibinfo {volume} {78}},\ \bibinfo
  {pages} {012352} (\bibinfo {year} {2008})}\BibitemShut {NoStop}%
\bibitem [{\citenamefont {Ising}(1925)}]{Ising25}%
  \BibitemOpen
  \bibfield  {author} {\bibinfo {author} {\bibfnamefont {E.}~\bibnamefont
  {Ising}},\ }\bibfield  {title} {\bibinfo {title} {Beitrag zur theorie des
  ferromagnetismus},\ }\href {https://doi.org/10.1007/BF02980577} {\bibfield
  {journal} {\bibinfo  {journal} {Zeitschrift f{\"u}r Physik}\ }\textbf
  {\bibinfo {volume} {31}},\ \bibinfo {pages} {253} (\bibinfo {year}
  {1925})}\BibitemShut {NoStop}%
\bibitem [{\citenamefont {Lucas}(2014)}]{lucas14}%
  \BibitemOpen
  \bibfield  {author} {\bibinfo {author} {\bibfnamefont {A.}~\bibnamefont
  {Lucas}},\ }\bibfield  {title} {\bibinfo {title} {Ising formulations of many
  np problems},\ }\bibfield  {journal} {\bibinfo  {journal} {Frontiers in
  Physics}\ }\textbf {\bibinfo {volume} {2}},\ \href
  {https://doi.org/10.3389/fphy.2014.00005} {10.3389/fphy.2014.00005} (\bibinfo
  {year} {2014})\BibitemShut {NoStop}%
\bibitem [{\citenamefont {Glover}\ \emph {et~al.}(2019)\citenamefont {Glover},
  \citenamefont {Kochenberger},\ and\ \citenamefont {Du}}]{glover19}%
  \BibitemOpen
  \bibfield  {author} {\bibinfo {author} {\bibfnamefont {F.}~\bibnamefont
  {Glover}}, \bibinfo {author} {\bibfnamefont {G.}~\bibnamefont
  {Kochenberger}},\ and\ \bibinfo {author} {\bibfnamefont {Y.}~\bibnamefont
  {Du}},\ }\bibfield  {title} {\bibinfo {title} {Quantum bridge analytics i: a
  tutorial on formulating and using qubo models},\ }\href
  {https://doi.org/10.1007/s10288-019-00424-y} {\bibfield  {journal} {\bibinfo
  {journal} {4OR}\ }\textbf {\bibinfo {volume} {17}},\ \bibinfo {pages} {335}
  (\bibinfo {year} {2019})}\BibitemShut {NoStop}%
\bibitem [{\citenamefont {Ebadi}\ \emph {et~al.}(2022)\citenamefont {Ebadi},
  \citenamefont {Keesling}, \citenamefont {Cain}, \citenamefont {Wang},
  \citenamefont {Levine}, \citenamefont {Bluvstein}, \citenamefont {Semeghini},
  \citenamefont {Omran}, \citenamefont {Liu}, \citenamefont {Samajdar},
  \citenamefont {Luo}, \citenamefont {Nash}, \citenamefont {Gao}, \citenamefont
  {Barak}, \citenamefont {Farhi}, \citenamefont {Sachdev}, \citenamefont
  {Gemelke}, \citenamefont {Zhou}, \citenamefont {Choi}, \citenamefont
  {Pichler}, \citenamefont {Wang}, \citenamefont {Greiner}, \citenamefont
  {Vuletić},\ and\ \citenamefont {Lukin}}]{Ebadi22}%
  \BibitemOpen
  \bibfield  {author} {\bibinfo {author} {\bibfnamefont {S.}~\bibnamefont
  {Ebadi}}, \bibinfo {author} {\bibfnamefont {A.}~\bibnamefont {Keesling}},
  \bibinfo {author} {\bibfnamefont {M.}~\bibnamefont {Cain}}, \bibinfo {author}
  {\bibfnamefont {T.~T.}\ \bibnamefont {Wang}}, \bibinfo {author}
  {\bibfnamefont {H.}~\bibnamefont {Levine}}, \bibinfo {author} {\bibfnamefont
  {D.}~\bibnamefont {Bluvstein}}, \bibinfo {author} {\bibfnamefont
  {G.}~\bibnamefont {Semeghini}}, \bibinfo {author} {\bibfnamefont
  {A.}~\bibnamefont {Omran}}, \bibinfo {author} {\bibfnamefont {J.-G.}\
  \bibnamefont {Liu}}, \bibinfo {author} {\bibfnamefont {R.}~\bibnamefont
  {Samajdar}}, \bibinfo {author} {\bibfnamefont {X.-Z.}\ \bibnamefont {Luo}},
  \bibinfo {author} {\bibfnamefont {B.}~\bibnamefont {Nash}}, \bibinfo {author}
  {\bibfnamefont {X.}~\bibnamefont {Gao}}, \bibinfo {author} {\bibfnamefont
  {B.}~\bibnamefont {Barak}}, \bibinfo {author} {\bibfnamefont
  {E.}~\bibnamefont {Farhi}}, \bibinfo {author} {\bibfnamefont
  {S.}~\bibnamefont {Sachdev}}, \bibinfo {author} {\bibfnamefont
  {N.}~\bibnamefont {Gemelke}}, \bibinfo {author} {\bibfnamefont
  {L.}~\bibnamefont {Zhou}}, \bibinfo {author} {\bibfnamefont {S.}~\bibnamefont
  {Choi}}, \bibinfo {author} {\bibfnamefont {H.}~\bibnamefont {Pichler}},
  \bibinfo {author} {\bibfnamefont {S.-T.}\ \bibnamefont {Wang}}, \bibinfo
  {author} {\bibfnamefont {M.}~\bibnamefont {Greiner}}, \bibinfo {author}
  {\bibfnamefont {V.}~\bibnamefont {Vuletić}},\ and\ \bibinfo {author}
  {\bibfnamefont {M.~D.}\ \bibnamefont {Lukin}},\ }\bibfield  {title} {\bibinfo
  {title} {Quantum optimization of maximum independent set using rydberg atom
  arrays},\ }\href {https://doi.org/10.1126/science.abo6587} {\bibfield
  {journal} {\bibinfo  {journal} {Science}\ }\textbf {\bibinfo {volume}
  {376}},\ \bibinfo {pages} {1209} (\bibinfo {year} {2022})}\BibitemShut
  {NoStop}%
\bibitem [{\citenamefont {Farhi}\ \emph {et~al.}(2014)\citenamefont {Farhi},
  \citenamefont {Goldstone},\ and\ \citenamefont {Gutmann}}]{farhi14}%
  \BibitemOpen
  \bibfield  {author} {\bibinfo {author} {\bibfnamefont {E.}~\bibnamefont
  {Farhi}}, \bibinfo {author} {\bibfnamefont {J.}~\bibnamefont {Goldstone}},\
  and\ \bibinfo {author} {\bibfnamefont {S.}~\bibnamefont {Gutmann}},\
  }\href@noop {} {\bibinfo {title} {A quantum approximate optimization
  algorithm}} (\bibinfo {year} {2014}),\ \Eprint
  {https://arxiv.org/abs/1411.4028} {arXiv:1411.4028 [quant-ph]} \BibitemShut
  {NoStop}%
\bibitem [{\citenamefont {Biamonte}\ \emph {et~al.}(2017)\citenamefont
  {Biamonte}, \citenamefont {Wittek}, \citenamefont {Pancotti}, \citenamefont
  {Rebentrost}, \citenamefont {Wiebe},\ and\ \citenamefont
  {Lloyd}}]{Biamonte17}%
  \BibitemOpen
  \bibfield  {author} {\bibinfo {author} {\bibfnamefont {J.}~\bibnamefont
  {Biamonte}}, \bibinfo {author} {\bibfnamefont {P.}~\bibnamefont {Wittek}},
  \bibinfo {author} {\bibfnamefont {N.}~\bibnamefont {Pancotti}}, \bibinfo
  {author} {\bibfnamefont {P.}~\bibnamefont {Rebentrost}}, \bibinfo {author}
  {\bibfnamefont {N.}~\bibnamefont {Wiebe}},\ and\ \bibinfo {author}
  {\bibfnamefont {S.}~\bibnamefont {Lloyd}},\ }\bibfield  {title} {\bibinfo
  {title} {Quantum machine learning},\ }\href
  {https://doi.org/10.1038/nature23474} {\bibfield  {journal} {\bibinfo
  {journal} {Nature}\ }\textbf {\bibinfo {volume} {549}},\ \bibinfo {pages}
  {195} (\bibinfo {year} {2017})}\BibitemShut {NoStop}%
\bibitem [{\citenamefont {Zhou}\ \emph {et~al.}(2020)\citenamefont {Zhou},
  \citenamefont {Wang}, \citenamefont {Choi}, \citenamefont {Pichler},\ and\
  \citenamefont {Lukin}}]{Zhou20}%
  \BibitemOpen
  \bibfield  {author} {\bibinfo {author} {\bibfnamefont {L.}~\bibnamefont
  {Zhou}}, \bibinfo {author} {\bibfnamefont {S.-T.}\ \bibnamefont {Wang}},
  \bibinfo {author} {\bibfnamefont {S.}~\bibnamefont {Choi}}, \bibinfo {author}
  {\bibfnamefont {H.}~\bibnamefont {Pichler}},\ and\ \bibinfo {author}
  {\bibfnamefont {M.~D.}\ \bibnamefont {Lukin}},\ }\bibfield  {title} {\bibinfo
  {title} {Quantum approximate optimization algorithm: Performance, mechanism,
  and implementation on near-term devices},\ }\href
  {https://doi.org/10.1103/PhysRevX.10.021067} {\bibfield  {journal} {\bibinfo
  {journal} {Phys. Rev. X}\ }\textbf {\bibinfo {volume} {10}},\ \bibinfo
  {pages} {021067} (\bibinfo {year} {2020})}\BibitemShut {NoStop}%
\bibitem [{\citenamefont {Feynman}(1982)}]{Feynman82}%
  \BibitemOpen
  \bibfield  {author} {\bibinfo {author} {\bibfnamefont {R.~P.}\ \bibnamefont
  {Feynman}},\ }\bibfield  {title} {\bibinfo {title} {Simulating physics with
  computers},\ }\href {https://doi.org/10.1007/BF02650179} {\bibfield
  {journal} {\bibinfo  {journal} {International Journal of Theoretical
  Physics}\ }\textbf {\bibinfo {volume} {21}},\ \bibinfo {pages} {467}
  (\bibinfo {year} {1982})}\BibitemShut {NoStop}%
\bibitem [{\citenamefont {White}(1992)}]{White92}%
  \BibitemOpen
  \bibfield  {author} {\bibinfo {author} {\bibfnamefont {S.~R.}\ \bibnamefont
  {White}},\ }\bibfield  {title} {\bibinfo {title} {Density matrix formulation
  for quantum renormalization groups},\ }\href
  {https://doi.org/10.1103/PhysRevLett.69.2863} {\bibfield  {journal} {\bibinfo
   {journal} {Phys. Rev. Lett.}\ }\textbf {\bibinfo {volume} {69}},\ \bibinfo
  {pages} {2863} (\bibinfo {year} {1992})}\BibitemShut {NoStop}%
\bibitem [{\citenamefont {White}(1993)}]{white93}%
  \BibitemOpen
  \bibfield  {author} {\bibinfo {author} {\bibfnamefont {S.~R.}\ \bibnamefont
  {White}},\ }\bibfield  {title} {\bibinfo {title} {Density-matrix algorithms
  for quantum renormalization groups},\ }\href
  {https://doi.org/10.1103/PhysRevB.48.10345} {\bibfield  {journal} {\bibinfo
  {journal} {Phys. Rev. B}\ }\textbf {\bibinfo {volume} {48}},\ \bibinfo
  {pages} {10345} (\bibinfo {year} {1993})}\BibitemShut {NoStop}%
\bibitem [{\citenamefont {Schollw\"ock}(2005)}]{DMRG}%
  \BibitemOpen
  \bibfield  {author} {\bibinfo {author} {\bibfnamefont {U.}~\bibnamefont
  {Schollw\"ock}},\ }\bibfield  {title} {\bibinfo {title} {The density-matrix
  renormalization group},\ }\href {https://doi.org/10.1103/RevModPhys.77.259}
  {\bibfield  {journal} {\bibinfo  {journal} {Rev. Mod. Phys.}\ }\textbf
  {\bibinfo {volume} {77}},\ \bibinfo {pages} {259} (\bibinfo {year}
  {2005})}\BibitemShut {NoStop}%
\bibitem [{\citenamefont {Or{\'u}s}(2014)}]{orus14}%
  \BibitemOpen
  \bibfield  {author} {\bibinfo {author} {\bibfnamefont {R.}~\bibnamefont
  {Or{\'u}s}},\ }\bibfield  {title} {\bibinfo {title} {A practical introduction
  to tensor networks: Matrix product states and projected entangled pair
  states},\ }\href {https://doi.org/https://doi.org/10.1016/j.aop.2014.06.013}
  {\bibfield  {journal} {\bibinfo  {journal} {Annals of Physics}\ }\textbf
  {\bibinfo {volume} {349}},\ \bibinfo {pages} {117} (\bibinfo {year}
  {2014})}\BibitemShut {NoStop}%
\bibitem [{\citenamefont {Or{\'u}s}(2019)}]{orus19}%
  \BibitemOpen
  \bibfield  {author} {\bibinfo {author} {\bibfnamefont {R.}~\bibnamefont
  {Or{\'u}s}},\ }\bibfield  {title} {\bibinfo {title} {Tensor networks for
  complex quantum systems},\ }\href {https://doi.org/10.1038/s42254-019-0086-7}
  {\bibfield  {journal} {\bibinfo  {journal} {Nature Reviews Physics}\ }\textbf
  {\bibinfo {volume} {1}},\ \bibinfo {pages} {538} (\bibinfo {year}
  {2019})}\BibitemShut {NoStop}%
\bibitem [{\citenamefont {Cirac}\ \emph {et~al.}(2021)\citenamefont {Cirac},
  \citenamefont {P\'erez-Garc\'{\i}a}, \citenamefont {Schuch},\ and\
  \citenamefont {Verstraete}}]{Cirac21}%
  \BibitemOpen
  \bibfield  {author} {\bibinfo {author} {\bibfnamefont {J.~I.}\ \bibnamefont
  {Cirac}}, \bibinfo {author} {\bibfnamefont {D.}~\bibnamefont
  {P\'erez-Garc\'{\i}a}}, \bibinfo {author} {\bibfnamefont {N.}~\bibnamefont
  {Schuch}},\ and\ \bibinfo {author} {\bibfnamefont {F.}~\bibnamefont
  {Verstraete}},\ }\bibfield  {title} {\bibinfo {title} {Matrix product states
  and projected entangled pair states: Concepts, symmetries, theorems},\ }\href
  {https://doi.org/10.1103/RevModPhys.93.045003} {\bibfield  {journal}
  {\bibinfo  {journal} {Rev. Mod. Phys.}\ }\textbf {\bibinfo {volume} {93}},\
  \bibinfo {pages} {045003} (\bibinfo {year} {2021})}\BibitemShut {NoStop}%
\bibitem [{\citenamefont {Ceperley}\ and\ \citenamefont
  {Alder}(1986)}]{Ceperley86}%
  \BibitemOpen
  \bibfield  {author} {\bibinfo {author} {\bibfnamefont {D.}~\bibnamefont
  {Ceperley}}\ and\ \bibinfo {author} {\bibfnamefont {B.}~\bibnamefont
  {Alder}},\ }\bibfield  {title} {\bibinfo {title} {Quantum monte carlo},\
  }\href {https://doi.org/10.1126/science.231.4738.555} {\bibfield  {journal}
  {\bibinfo  {journal} {Science}\ }\textbf {\bibinfo {volume} {231}},\ \bibinfo
  {pages} {555} (\bibinfo {year} {1986})}\BibitemShut {NoStop}%
\bibitem [{\citenamefont {Austin}\ \emph {et~al.}(2012)\citenamefont {Austin},
  \citenamefont {Zubarev},\ and\ \citenamefont {Lester}}]{Austin12}%
  \BibitemOpen
  \bibfield  {author} {\bibinfo {author} {\bibfnamefont {B.~M.}\ \bibnamefont
  {Austin}}, \bibinfo {author} {\bibfnamefont {D.~Y.}\ \bibnamefont
  {Zubarev}},\ and\ \bibinfo {author} {\bibfnamefont {W.~A.~J.}\ \bibnamefont
  {Lester}},\ }\bibfield  {title} {\bibinfo {title} {Quantum monte carlo and
  related approaches},\ }\href {https://doi.org/10.1021/cr2001564} {\bibfield
  {journal} {\bibinfo  {journal} {Chemical Reviews}\ }\textbf {\bibinfo
  {volume} {112}},\ \bibinfo {pages} {263} (\bibinfo {year}
  {2012})}\BibitemShut {NoStop}%
\bibitem [{\citenamefont {Zwolak}\ and\ \citenamefont
  {Vidal}(2004)}]{zwolak94}%
  \BibitemOpen
  \bibfield  {author} {\bibinfo {author} {\bibfnamefont {M.}~\bibnamefont
  {Zwolak}}\ and\ \bibinfo {author} {\bibfnamefont {G.}~\bibnamefont {Vidal}},\
  }\bibfield  {title} {\bibinfo {title} {Mixed-state dynamics in
  one-dimensional quantum lattice systems: A time-dependent superoperator
  renormalization algorithm},\ }\href
  {https://doi.org/10.1103/PhysRevLett.93.207205} {\bibfield  {journal}
  {\bibinfo  {journal} {Phys. Rev. Lett.}\ }\textbf {\bibinfo {volume} {93}},\
  \bibinfo {pages} {207205} (\bibinfo {year} {2004})}\BibitemShut {NoStop}%
\bibitem [{\citenamefont {Daley}\ \emph {et~al.}(2004)\citenamefont {Daley},
  \citenamefont {Kollath}, \citenamefont {Schollwöck},\ and\ \citenamefont
  {Vidal}}]{Daley04}%
  \BibitemOpen
  \bibfield  {author} {\bibinfo {author} {\bibfnamefont {A.~J.}\ \bibnamefont
  {Daley}}, \bibinfo {author} {\bibfnamefont {C.}~\bibnamefont {Kollath}},
  \bibinfo {author} {\bibfnamefont {U.}~\bibnamefont {Schollwöck}},\ and\
  \bibinfo {author} {\bibfnamefont {G.}~\bibnamefont {Vidal}},\ }\bibfield
  {title} {\bibinfo {title} {Time-dependent density-matrix
  renormalization-group using adaptive effective hilbert spaces},\ }\href
  {https://doi.org/10.1088/1742-5468/2004/04/P04005} {\bibfield  {journal}
  {\bibinfo  {journal} {Journal of Statistical Mechanics: Theory and
  Experiment}\ }\textbf {\bibinfo {volume} {2004}},\ \bibinfo {pages} {P04005}
  (\bibinfo {year} {2004})}\BibitemShut {NoStop}%
\bibitem [{\citenamefont {Verstraete}\ \emph {et~al.}(2004)\citenamefont
  {Verstraete}, \citenamefont {Garc\'{\i}a-Ripoll},\ and\ \citenamefont
  {Cirac}}]{verstraete04}%
  \BibitemOpen
  \bibfield  {author} {\bibinfo {author} {\bibfnamefont {F.}~\bibnamefont
  {Verstraete}}, \bibinfo {author} {\bibfnamefont {J.~J.}\ \bibnamefont
  {Garc\'{\i}a-Ripoll}},\ and\ \bibinfo {author} {\bibfnamefont {J.~I.}\
  \bibnamefont {Cirac}},\ }\bibfield  {title} {\bibinfo {title} {Matrix product
  density operators: Simulation of finite-temperature and dissipative
  systems},\ }\href {https://doi.org/10.1103/PhysRevLett.93.207204} {\bibfield
  {journal} {\bibinfo  {journal} {Phys. Rev. Lett.}\ }\textbf {\bibinfo
  {volume} {93}},\ \bibinfo {pages} {207204} (\bibinfo {year}
  {2004})}\BibitemShut {NoStop}%
\bibitem [{\citenamefont {Manzano}(2020)}]{Manzano20}%
  \BibitemOpen
  \bibfield  {author} {\bibinfo {author} {\bibfnamefont {D.}~\bibnamefont
  {Manzano}},\ }\bibfield  {title} {\bibinfo {title} {{A short introduction to
  the Lindblad master equation}},\ }\href {https://doi.org/10.1063/1.5115323}
  {\bibfield  {journal} {\bibinfo  {journal} {AIP Advances}\ }\textbf {\bibinfo
  {volume} {10}},\ \bibinfo {pages} {025106} (\bibinfo {year}
  {2020})}\BibitemShut {NoStop}%
\bibitem [{Note1()}]{Note1}%
  \BibitemOpen
  \bibinfo {note} {Here we focus on measurements of the first kind, comp.~\cite
  {gardiner00}}\BibitemShut {NoStop}%
\bibitem [{\citenamefont {Willsch}\ \emph {et~al.}(2022)\citenamefont
  {Willsch}, \citenamefont {Willsch}, \citenamefont {Gonzalez~Calaza},
  \citenamefont {Jin}, \citenamefont {De~Raedt}, \citenamefont {Svensson},\
  and\ \citenamefont {Michielsen}}]{willsch22}%
  \BibitemOpen
  \bibfield  {author} {\bibinfo {author} {\bibfnamefont {D.}~\bibnamefont
  {Willsch}}, \bibinfo {author} {\bibfnamefont {M.}~\bibnamefont {Willsch}},
  \bibinfo {author} {\bibfnamefont {C.~D.}\ \bibnamefont {Gonzalez~Calaza}},
  \bibinfo {author} {\bibfnamefont {F.}~\bibnamefont {Jin}}, \bibinfo {author}
  {\bibfnamefont {H.}~\bibnamefont {De~Raedt}}, \bibinfo {author}
  {\bibfnamefont {M.}~\bibnamefont {Svensson}},\ and\ \bibinfo {author}
  {\bibfnamefont {K.}~\bibnamefont {Michielsen}},\ }\bibfield  {title}
  {\bibinfo {title} {Benchmarking advantage and d-wave 2000q quantum annealers
  with exact cover problems},\ }\href
  {https://doi.org/10.1007/s11128-022-03476-y} {\bibfield  {journal} {\bibinfo
  {journal} {Quantum Information Processing}\ }\textbf {\bibinfo {volume}
  {21}},\ \bibinfo {pages} {141} (\bibinfo {year} {2022})}\BibitemShut
  {NoStop}%
\bibitem [{\citenamefont {Gardiner}\ and\ \citenamefont
  {Zoller}(2000)}]{gardiner00}%
  \BibitemOpen
  \bibfield  {author} {\bibinfo {author} {\bibfnamefont {C.~W.}\ \bibnamefont
  {Gardiner}}\ and\ \bibinfo {author} {\bibfnamefont {P.}~\bibnamefont
  {Zoller}},\ }\href@noop {} {\emph {\bibinfo {title} {Quantum Noise}}},\
  \bibinfo {edition} {2nd}\ ed.,\ edited by\ \bibinfo {editor} {\bibfnamefont
  {H.}~\bibnamefont {Haken}}\ (\bibinfo  {publisher} {Springer},\ \bibinfo
  {year} {2000})\BibitemShut {NoStop}%
\end{thebibliography}%

\appendix

\section{The Algorithm}\label{app-algo}

Here we present two variants of the algorithmic structure of implementing the Lindblad master equation in the ORQA formalism.
These algorithms assume a data structure of $\rho$ that is dynamically sized and are shown as Algorithm 1 and Algorithm 2.
As discussed in the main-text, it is also possible to consider a trivial map, i.e.\ directly associating the multi-indices $I$ with physical memory addresses.
That method is faster but requires exponentially large memory.
While these two variants outline a straight-forward approach to perform an integration step in the ORQA formalism, 
we emphasize that they can be utilized in various integrators, and do not imply an Euler-method. 
For instance for the results in the main-text, we have implemented these update rules in a $4$th-order Runge-Kutta method. 

In order to understand the instantiation of the numerical implementation of $\rho$ and $H$, consider how to obtain multi-indices.
Consider the multi-index of a local Pauli matrix, e.g. $\sigma_I = \sigma_k^j$ which is the $k$th Pauli matrix at the $j$th position.
$k$ has a two-bit representation, such that 
\begin{equation}
    I = k \ll (2j),
\end{equation}
where $\ll$ denotes the bit-shift operation.
For instance in the case of $\sigma_z^4$ it is $j=4$ and $k=3=11_2$.
Therefore 
\begin{equation}
    0000000011_2 \ll (2\times 4) \rightarrow 1100000000_2 = 3072.
\end{equation}
Hence, the mutli-index for $\sigma_z^4=\sigma_I$ is $I=3072$. 
Note that counting two-level systems starts at $0$.
From this we construct many-body multi-indices using the bit-wise or operation $I \lor J$.
Consider the operator $\sigma_K = \sigma_x\otimes\sigma_y$, where we want to identify $K$.
From the construction above we obtain $\sigma_I=\sigma_x\otimes\sigma_0$ and $\sigma_J=\sigma_0\otimes\sigma_y$, and then $K=I \lor J$.

\begin{algorithm}[t]
    \caption{Variant 1 of the Lindblad master equation in ORQA}
    Initialize map for initial state $\rho$\;
    Initialize map for Hamiltonian $H$\;
    Initialize empty map for $\dot\rho$\;
    Initialize truncation threshold $\epsilon\geq 0$\;
    Initialize coefficients $\gamma_-\geq 0$,$\gamma_z\geq 0$\;
    Initialize time discretization $\Delta t\geq 0$\;
    \For{time-steps t}{
        Clear $\dot\rho$\;
        \For{$\{I,\rho_I\}$ in $\rho$}{
            \For{$\{J,H_J\}$ in $H$}{
                $\dot\rho_{I\veebar J}\mathrel{+}= 2 Y^{I,J} \rho_I H_J$\;
            }
            $\dot\rho_I\mathrel{-}= \gamma_z 2\rho_I \sum_j I_{j,1}\veebar I_{j,2}$\;
            $\dot\rho_I\mathrel{-}= \gamma_- (\sum_j \delta_{I_j,3}\rho_{I\veebar 3 \times 4^j}+\frac{\oplus}{2}\rho_I)$\;
        }
        \For{$\{I,\dot\rho_I\}$ in $\dot\rho$}{
            $\rho_I\mathrel{+}=\Delta t \dot\rho_I$\;
            \If{$|\rho_I|\leq \epsilon$}{
                Delete $\{I,\rho_I\}$ from $\rho$\;
            }
        }
    }
\end{algorithm}

\begin{algorithm}[t]
    \caption{Variant 2 of the Lindblad master equation in ORQA}
    Initialize map for initial state $\rho$\;
    Initialize map for Hamiltonian $H$\;
    Initialize empty map for $\dot\rho$\;
    Initialize truncation threshold $\epsilon\geq 0$\;
    Initialize coefficients $\gamma_- \geq 0$,$\gamma_z \geq 0$\;
    Initialize time discretization $\Delta t \geq 0$\;
    \ForAll{time-steps t}{
        Clear $\dot\rho$\;
        \ForAll{$\{J,H_J\}$ in $H$}{
            \ForAll{$\{I,\rho_I\}$ in $\rho$}{
                insert $\{I \veebar J, 0\}$ in $\dot\rho$\;
            }
        }
        \ForAll{$\{I,\dot\rho_I\}$ in $\dot\rho$}{
            \ForAll{$\{J,H_J\}$ in $H$}{
                $\dot\rho_I\mathrel{+}= 2 Y^{I\veebar J,J} \rho_{I\veebar J}H_J$\;
                $\dot\rho_I\mathrel{-}= \gamma_z 2\rho_I \sum_j I_{j,1}\veebar I_{j,2}$\;
                $\dot\rho_I\mathrel{-}= \gamma_- (\sum_j \delta_{I_j,3}\rho_{I\veebar 3 \times 4^j}+\frac{\oplus}{2}\rho_I)$\;
            }
        }
        \ForAll{$\{I,\dot\rho_I\}$ in $\dot\rho$}{
            $\rho_I\mathrel{+}=\Delta t \dot\rho_I$\;
            \If{$|\rho_I|\leq \epsilon$}{
                Delete $\{I,\rho_I\}$ from $\rho$\;
            }
        }
    }
\end{algorithm}

As a minimal example, the Hamiltonian $H=a\sigma_x^0+b\sigma_x^1+c \sigma_z^1\otimes\sigma_z^2$ is represented as a map of the form
\begin{align}
    H &= \{\{0001_2,a\},\{0100_2,b\},\{1111_2,c\}\} \\
    &= \{\{1,a\},\{4,b\},\{15,c\}\}.
\end{align}

In the case of a dynamically sized map, it is necessary to identify which new elements of $\rho$ might occur during time-evolution.
In Algorithm 1, we circumvent this by looping over $\rho$ and $H$ and accessing the elements $I\veebar J$.
In Algorithm 2, we include a preliminary loop that prepares all possible $I\veebar J$ in $\dot\rho$.
In practice, it is beneficial to introduce extra intermediate steps, such as comparing the magnitude of components $\dot\rho_I$ with the truncation threshold as well in order to avoid inserting components into $\rho$ just to immediately delete them again, due to the dynamical truncation.

\section{Partial trace}\label{app-partial}

A useful operation in the context of numerical quantum dynamics is the partial trace.
In the ORQA formalism, this is very straight-forward. 
Tracing out the $j$th two-level system is given by
\begin{align*}
    \rho\rightarrow\rho'&=\mathrm{Tr}_j(\sum_I \rho_I \sigma_I) = \sum_I \rho_I \mathrm{Tr}_j(\sigma_I) \\ &= \sum_I \rho_I \delta_{I_j,0} \otimes_{i\not=j}\sigma_{I_i},
\end{align*}
which amounts to dropping all terms of multi-indices $I$ of which the $j$th two-bit string $I_j$ is not zero. 
In principle the $n$-body multi-indices could be recast into $n-1$-bit ones, which amounts to the reduction of the Hilbert space that accompanies the partial trace.
However, this operation is not necessary in general and in the ORQA formalism the data-structure is not affected by this, which is very convenient for dynamics with intermediate partial trace operations.

\section{Symmetrization}\label{symmetry}

Reconsider the von Neumann equation as written in the main-text
\begin{equation}
    \dot\rho_I = 2{\sum_{J}}^\prime \rho_{I\veebar J} H_J Y^{I\veebar J,J}.
\end{equation}
In systems that display symmetries, the Hamiltonian and the density operator may be invariant under a set of transformations $S$, e.g.\ permutations of spins. 
This in particular means in the ORQA formalism, that $\rho_I=\rho_{s(I)}$ and $H_I=H_{s(I)}$, $\forall s\in S$. 
This implies an equivalence class on the space of multi-indices with $I \sim J$, iff $\exists s\in S$ such that $I=s(J)$.
It is therefore sufficient to only keep track of a unique representative of each equivalence class under $S$, which has the potential to significantly increase the efficiency of the numerics.
Since in the ORQA formalism we are dealing with multi-indices, a natural choice for such representatives is the smallest number within an equivalence class.
We write
\begin{equation}
    I_S = \min_{s\in S} s(I).
\end{equation}

Therefore, the von Neumann equation becomes
\begin{align}
    \dot\rho_{I_S} 
    &= 2 {\sum_{J}}^\prime \rho_{(I\veebar J)_S} H_{J} Y^{I \veebar J, J},
\end{align}
where we used that $H_{s(J)}=H_J$, $\forall s\in S$. 
Note that this calculation only necessitates the representative $(I\veebar J)_S$, but not any corresponding element of the symmetry group.
The efficiency of this relies on how difficult it is to identify $(I\veebar J)_S$ from $I$, which depends on the nature of $S$ itself.

Consider the group of arbitrary permutations of spins which we denote as $S_P$. 
$S_P$ is a large symmetry group, in which it is easy to identify minimal elements.
One simply reorders all 2-bit strings to be sorted, e.g.
\begin{equation}
    I_{S_2} = \underbrace{00|00|}_{n_{I,00_2}}\dots\underbrace{|01|}_{n_{I,01_2}}\dots\underbrace{|10|}_{n_{I,10_2}}\dots\underbrace{|11|11}_{n_{I,11_2}},
\end{equation}
where $n_{I,j}$ is the amount of two-bit strings with value $j$ in $I$.
This group of permutations separates the $4^{n}$ multi-indices into $n_P$ equivalence classes of varying sizes.
These equivalence classes are characterized by how many of each local index appear. 
This means they are characterized by all possible combinations of $\{n_0,n_1,n_2,n_3\}$ such that $n_0+n_1+n_2+n_3=n$.
This amounts to $n_P = (n+1)(n+2)(n+3)/6$ combinations, i.e.\ $\mathcal{O}(n^3)$ many.
This is a severe reduction in complexity of systems that display this type of symmetry.

\section{Derivation of Lindblad Terms}\label{app-lindblad}

The Lindblad action in Eq.~\ref{lindgeneral} of a single local $\sigma_z$ term, i.e.\ $z=3\times 4^j$, evaluates to
\begin{align}
    \mathcal{L}_z[\rho] 
    &= \sum_{I=1}^{4^n-1} \rho_I (\sigma_z \sigma_I \sigma_z -  \sigma_I)\\
    &= \sum_{I=1}^{4^n-1} \rho_I (i^{B(I,z)+B(z,I\veebar z)} \sigma_I - \sigma_I)\\
    &= -2 \sum_{I=1}^{4^n-1} \rho_I \sigma_I (\delta_{I_j,x}+\delta_{I_j,y})
\end{align}
such that
\begin{equation}
    \mathcal{L}_{z,I}[\rho] = \begin{cases}
        0, &I_j = 0,z\\
        -2\rho_I, &I_j = x,y
    \end{cases}.
\end{equation}

Therefore, in the case that all $n$ local $\sigma_z$ terms act as dissipative Lindblad operators, the full action becomes
\begin{align}
    (\sum_{j=0}^{n-1} \mathcal{L}_{z_j}[\rho])_I &= -2\rho_I \sum_{j=0}^{n-1} I_{j,1}\veebar I_{j,2}.
\end{align}
Here we have used that for the two-bit string $I_j=a|b$, with $a,b\in\{0,1\}$, it is $(a|b=10_2 \lor a|b=01_2)\iff a\veebar b$. 

The Lindblad action for single local $\sigma_\pm=(\sigma_x\pm i \sigma_y)/2$ terms evaluates to
\onecolumngrid  
\vspace{0.5cm}
\rule{8.5cm}{0.01cm}
\begin{align} 
    \mathcal{L}^j_\pm[\rho] 
    &= \frac{1}{4}\sum_{I=0}^{4^n-1} \rho_I ((\sigma_x\pm i\sigma_y) \sigma_I (\sigma_x\mp i\sigma_y) - \frac{1}{2}\{(\sigma_x\mp i\sigma_y)(\sigma_x\pm i\sigma_y),\sigma_I\})\\
    &= \frac{1}{4}\sum_{I=0}^{4^n-1} \rho_I (\sigma_x\sigma_I\sigma_x\pm i\sigma_y\sigma_I\sigma_x \mp i \sigma_x\sigma_I\sigma_y + \sigma_y\sigma_I\sigma_y - 2\sigma_I \pm \{\sigma_z,\sigma_I\})\\
    &= \frac{1}{4}\sum_{I=0}^{4^n-1} \rho_I (2(\delta_{I_j,0}-\delta_{I_j,3})\sigma_I \pm 2(\delta_{I_j,0}-\delta_{I_j,3})\sigma_{I\veebar z}  - 2\sigma_I \pm 2\sigma_{I \veebar z}(\delta_{I_j,0} + \delta_{I_j,3}) )\\
    &= \frac{1}{4}\sum_{I=0}^{4^n-1} \rho_I ((2\delta_{I_j,0}-2\delta_{I_j,3}-2)\sigma_I \pm 4 \delta_{I_j,0} \sigma_{I\veebar z})\\
    {\mathcal{L}^j_\pm[\rho]}_I &= -\frac{1}{2}(\delta_{I_j,1}+\delta_{I_j,2}+2\delta_{I_j,3})\rho_I \pm \delta_{I_j,3}\rho_{I \veebar z}
    = -\frac{1}{2}(I_{j,1}+I_{j,2})\rho_I \pm \delta_{I_j,3}\rho_{I \veebar z}
\end{align}
\hspace{9.25cm}\rule{8.5cm}{0.01cm}
\twocolumngrid
Here we again used the notation of single-body multi-indices $x=1\times 4^j$, $y=2\times 4^j$, and $z=3\times 4^j$.
The presence of this type of dissipation in all the two-level systems amounts to the total Lindblad action
\begin{align}
    (\sum_{j=0}^{n-1}\mathcal{L}^j_{\pm}[\rho])_I = \pm \sum_{j=0}^{n-1} \delta_{I_j,3} \rho_{I\mathrm{xor}3\times 4^j} - \rho_I \frac{\oplus I}{2}. 
\end{align}
Here $\oplus$ denotes the binary digit sum, which is a noteworthy analytical curiosity.

\section{MIS Ground States}\label{app-groundstate}

To confirm that the degenerate ground state of $H_\mathrm{MIS}$ in Eq.~\ref{HMIS} contains all and only MISs, consider the change in energy due to spin-flips.
Flipping a spin provides a change in energy equal to 
\begin{equation}
    \Delta E = \pm(n_\uparrow - \frac{1}{2}),
\end{equation}
where the positive sign is associated with flipping a down-spin up and the negative sign with flipping an up-spin down.
$n_\uparrow$ is the number of neighbors of the flipped spin that are in the up-state.
Therefore, energy can be reduced through a spin-flip unless for all down-spins it is $n_\uparrow > 0$ and for all up-spins it is $n_\uparrow = 0$.
This means independent sets always gain energy from single spin-flips, and for every non-independent set state there is at least one spin-flip available that reduces the energy.
Strictly, this only shows that the ground state consists of independent sets.
However, any independent set of size $m$ has energy 
\begin{equation}
    E=-\frac{m}{2}.
\end{equation}
Therefore, the degenerate ground state of $H_\mathrm{MIS}$ always consists of exactly all MISs of $G$.

\end{document}